# Non-slipping adhesive contact of rigid cylinder on an elastic power-law graded half-space

Fan Jin    Xu Guo[1]

*State Key Laboratory of Structural Analysis for Industrial Equipment*
*Department of Engineering Mechanics*
*Dalian University of Technology, Dalian, 116023, P.R. China*

**Abstract**

In this paper, adhesive contact of a rigid cylinder on an elastic power-law graded half-space is studied analytically with the theory of weakly singular integral equation and orthogonal polynomial method. Emphasis is placed on the coupling effect between tangential and normal directions which was often neglected in previous works. Our analysis shows that the coupling effect tends to minimize the contact area in the compressive regime. The effect of bending moment on the adhesion behavior is also examined. Like a pull-off force, there also exist a critical bending moment at which the cylinder can be bended apart from the substrate. However, unlike pull-off force, the critical bending moment is insensitive to the gradient exponent of the graded material.

---

[1] Corresponding author.    Tel: +86-411-84707807    E-mail address: guoxu@dlut.edu.cn

# 1. Introduction

The adhesion forces between surfaces which are induced by the intermolecular interactions may significantly affect the contact behavior of small size objects. Such situation often arises in micro-electro-mechanical-systems (MEMS), scanning probe microscope (SPM) measurements and micro/ nano- tribology applications. Therefore, the study of adhesive contact behavior has received increasing attention in recent years (Johnson et al., 1971; Derjaguin et al., 1975; Maugis, 1992; Barquins, 1988; Chaudhury et al., 1996; Baney and Hui, 1997; Greenwood, 1997; Greenwood and Johnson, 1998; Barthel, 1998; Hui et al., 2001; Chen and Gao 2006, Chen and Gao 2007). It is worth noting, however, that only homogeneous elastic materials are considered in all these works.

Problems arise when the contact behavior of non-homogeneous materials is taken into account. Mathematical complexity involved in this kind of problem often leads to great difficulties in obtaining the exact stress and displacement fields in closed form. For instance, the singular integral equations for contact analysis of two-dimensional power-law graded half-space are not of Cauchy but Abel type, whose solution requires more advanced techniques.

Previous studies on the contact behavior for graded materials can be found in the works of Booker et al. (1985a, 1985b), Gibson (1967), Gibson et al. (1971,1975), Brown and Gibson (1972), Awojobi and Gibson (1973), Calladine and Greenwood (1978). In most of these literatures, only non-adhesive contact problems under normal forces with specific Poisson's ratio and gradient exponents were considered. Giannakopoulos et al.

made a systematic exploration into the micromechanics of indentation on graded elastic solids (Giannakopoulos and Suresh, 1997a, b). In their work, the force-depth relations, the depth-contact radius relations and stress/displacement fields at the contact surface were examined through a combination of analytical, computational and experimental investigations. Furthermore, two-dimensional contact problem of power-law graded materials was also addressed (Giannakopoulos and Pallot, 2000). Chen et al. (2009a, b) extended these results to examine the adhesive behavior of graded materials. By using the superposition technique proposed by Johnson (1985) and Maugis (1992), they obtained the analytical solutions for pull-off force and critical contact radius of the contact area between a rigid sphere and a graded elastic half-space. The same technique is also employed to study the corresponding plane strain problem. In most of the above mentioned works, however, the tangential traction inside the contact region was either ignored or assumed to be independent of the normal traction. In addition, the possible bending moment that may be induced by the external force was also totally neglected in these works.

The main purpose of the present paper is to elucidate the coupling effect between normal and tangential tractions on the adhesive behavior of power-law graded materials from a theoretical point of view. To this end, the orthogonal polynomial method developed by Popov (1992) will be employed to find the exact tangential and normal tractions inside the contact region for rigid cylinder punch taking into account the coupling effect. The rest of the paper is organized as follows: In Section 2, the contact problem under consideration and its solution approach are introduced. According to whether the bending effect is included or not, the adhesive contact problem under different boundary conditions will be examined analytically in Section 3. Analysis results and discussions are presented in Section 4. The paper is ended with some concluding remarks in the final section.

## 2. Contact problem and the solution approach

The problem under consideration is the same as that discussed in Giannakopoulos and Pallot (2000) and Chen et al. (2009a). As shown in Fig.1, a rigid cylinder of radius $R$ is in non-slipping adhesive contact with an elastically graded half-space with constant Poisson ratio $\mu$ and elastic modulus varying with depth according to a power law. It is also assumed that the width of the contact region is $2a$, which is symmetric with respect to the $z$-axis. As shown in Guo and Jin (2009), this assumption is reasonable especially when the contact behavior before and at pull-off is focused on. In the present work, two loading conditions are contained: (a) Only the force effect is considered; (b) Both the effects of force and bending moment are considered.

The Young's modulus of the graded half-space varies in the following form

$$E = E_0(z/c_0)^v, \qquad 0 < v < 1, \qquad (2.1)$$

where $E_0$ is a reference modulus, $c_0$ a characteristic depth ($c_0 > 0$) and $v$ is the gradient exponent. The Poisson's ratio is assumed to be constant along the depth.

With use of the surface Green's function, the interfacial displacements $\bar{u}_x(x)$ and $\bar{u}_y(y)$ of the half-space can be related to the interfacial normal and tangential tractions $p(x)$ and $q(x)$ as follows (Popov, 1972; Giannakopoulos and Pallot, 2000)

$$\theta_0 \int_{-a}^{a} \frac{p(y)dy}{v|x-y|^v} + \theta_1 \int_{-a}^{a} \frac{\text{sign}(x-y)}{|x-y|^v} q(y)dy = \bar{u}_y(x), \tag{2.2a}$$

$$\theta_2 \int_{-a}^{a} \frac{q(y)dy}{v|x-y|^v} - \theta_3 \int_{-a}^{a} \frac{\text{sign}(x-y)}{|x-y|^v} p(y)dy = \bar{u}_x(x), \tag{2.2b}$$

where

$$\theta_0 = \frac{(1-\mu^2)\beta c_0^v C_v \sin\frac{\beta\pi}{2}}{(1+v)E_0}, \qquad \theta_2 = \frac{(1-\mu^2)(1+v)c_0^v C_v \sin\frac{\beta\pi}{2}}{\beta E_0},$$

$$\theta_3 = \theta_1 = -\frac{(1-\mu^2)c_0^v C_v \cos\frac{\beta\pi}{2}}{vE_0},$$

$$C_v = \frac{2^{1+v}\Gamma\left(\frac{3+v+\beta}{2}\right)\Gamma\left(\frac{3+v-\beta}{2}\right)}{\pi\Gamma(2+v)}, \qquad \beta = \sqrt{(1+v)\left(1-\frac{\mu v}{1-\mu}\right)} \tag{2.3}$$

and $\Gamma = \Gamma(x)$ is the Gamma function.

Under given interfacial displacements $\bar{u}_x(x)$ and $\bar{u}_y(x)$, Eq. (2.2) is a system of coupled Abel type singular integral equations for unknown interfacial tractions. From Eq. (2.2), it can be seen that coupling exists between the normal and tangential direction problems: the tangential interfacial traction can induce normal displacement whilst the normal interfacial traction can also induce tangential displacement. Theoretically, the coupling effect can only be neglected when $\theta_1 = \theta_3 = 0$. In previous studies (Giannakopoulos and Pallot, 2000; Chen et al., 2009a), this coupling effect has been neglected. Up to now, it is still unclear whether the coupling effect will play an important role on the adhesive contact behavior of graded materials or not. In the present work, this issue will be addressed analytically by employing the seminal results developed by Popov (1972) for solving the weakly singular integral equation.

Let us first rewrite Eq. (2.2) in dimensionless form. This can be achieved by letting

$$x = a\xi, \quad y = a\eta, \quad |\xi| \leq 1, \quad |\eta| \leq 1. \tag{2.4}$$

Substituting Eq. (2.4) into Eq. (2.2) and after some algebraic manipulations, we have

$$a^{1-v}\left[\frac{\theta_0}{\theta_1}\int_{-1}^{1}\frac{p(a\eta)d\eta}{v|\xi-\eta|^v} + \int_{-1}^{1}\frac{\text{sign}(\xi-\eta)}{|\xi-\eta|^v}q(a\eta)d\eta\right] = \frac{1}{\theta_1}\bar{u}_y(a\xi), \tag{2.5a}$$

$$a^{1-v}\left[\frac{\theta_2}{\theta_3}\int_{-1}^{1}\frac{q(a\eta)d\eta}{v|\xi-\eta|^v} - \int_{-1}^{1}\frac{\text{sign}(\xi-\eta)}{|\xi-\eta|^v}p(a\eta)d\eta\right] = \frac{1}{\theta_3}\bar{u}_x(a\xi). \tag{2.5b}$$

By introducing

$$\kappa = \sqrt{\theta_0\theta_3/(\theta_1\theta_2)}, \tag{2.6}$$

Eq. (2.5) can be further transformed into the following complex form:

$$\int_{-1}^{1}\frac{\text{sign}(\xi-\eta) + i\cot(\lambda\pi/2)}{|\xi-\eta|^v}\chi(\eta)d\eta = g(\xi), \tag{2.7}$$

where

$$\chi(\eta) = r(\eta) + is(\eta), \tag{2.8a}$$

$$r(\eta) = \kappa^{\frac{1}{2}}a^{1-v}p(a\eta), \quad s(\eta) = \kappa^{-\frac{1}{2}}a^{1-v}q(a\eta), \tag{2.8b}$$

$$\cot(\lambda\pi/2) = \sqrt{\theta_0\theta_3/(\theta_1\theta_2)}/v \tag{2.8c}$$

and

$$g(\xi) = i\kappa^{-\frac{1}{2}}(\theta_1)^{-1}\bar{u}_y(a\xi) - \kappa^{\frac{1}{2}}(\theta_3)^{-1}\bar{u}_x(a\xi). \tag{2.8d}$$

According to Popov (1972), the exact solution of the integral equation in (2.7) can be expressed as a series of Jacobi polynomials, that is

$$\chi(\xi) = \sum_{m=0}^{\infty}\frac{g_m P_m^\rho(\xi)}{i\sigma_v\lambda_m\psi_\rho(\xi)}, \tag{2.9}$$

where $P_m^\rho(\xi) = P_m^{-w-i\rho,-w+i\rho}(\xi)$ is the Jacobi polynomial of order $m$ with index $(-w - i\rho, -w + i\rho)$ and

$$\psi_\rho(\xi) = \frac{(1-\xi)^{w+i\rho}}{(1-\xi)^{-w+i\rho}}, \quad w = \frac{1-v}{2}, \quad \rho = \frac{1}{2\pi}\ln\frac{\sin[(v+\lambda)\pi/2]}{\sin[(v-\lambda)\pi/2]}, \tag{2.10}$$

$$\sigma_v = \frac{2\pi\sqrt{\sin\left[(v+\lambda)\pi/2\right]\sin\left[(v-\lambda)\pi/2\right]}}{\sin(v\pi)\sin(\lambda\pi/2)}, \quad \lambda = \beta - 1, \tag{2.11}$$

$$g_m = \int_{-1}^{1} \frac{g(\xi) P_m^{-\rho}(\xi)}{\psi_{-\rho}(\xi)} d\xi, \quad \lambda_m = \frac{2^v |\Gamma(1-w+i\rho+m)|^2}{m!^2 \, (v+2m)\Gamma(v)}. \tag{2.12}$$

Once $\chi(\eta)$ is determined, the interfacial tractions $p(a\eta)$ and $q(a\eta)$ can be obtained from its real and imaginary part, respectively. Furthermore, the width of the contact area can be calculated from

$$\left.\frac{\partial U_T}{\partial a}\right|_{\bar{u}_x(x),\bar{u}_y(x)} = \left.\frac{\partial (U_E + U_S)}{\partial a}\right|_{\bar{u}_x(x),\bar{u}_y(x)} = 0, \tag{2.13}$$

where the derivative is taken under fixed $\bar{u}_x(x)$ and $\bar{u}_y(x)$. In Eq. (2.13), $U_E$ denotes the elastic strain energy stored in the grated half-space and $U_S = -2a\Delta\gamma$ is associated with the surface energy due to the work of adhesion $\Delta\gamma$.

In the following sections, the above solution procedures will be employed to investigate the adhesive contact behavior of rigid cylinder indenter-graded half-space system under different loading conditions with coupling effect taken into account.

## 3. Analysis

### 3.1 Contact behavior without bending effect

In this case, it is assumed that only the force applied on the rigid cylinder is considered and the effect of bending moment is totally omitted. As a consequence, the rigid cylinder just experiences translation. This assumption has been adopted by many authors in the analysis of adhesive contact behavior by supposing that the external force is properly applied.

Under this circumstance, the tangential and normal displacements along the interface of the elastic graded half-space can be expressed as

$$\bar{u}_x = \varepsilon, \qquad \bar{u}_y = h - \frac{x^2}{2R}, \qquad |x| \leq a, \tag{3.1}$$

where $\varepsilon$ and $h$ are two constants.

Replacing $x$ with $a\xi$ in Eq. (3.1) and inserting it to Eq. (2.8d) leads to

$$g(\xi) = -\frac{\mathrm{i}}{R_1}\xi^2 + (\mathrm{i}h^* - \varepsilon^*), \tag{3.2}$$

where

$$h^* = \kappa^{-\frac{1}{2}}(\theta_1)^{-1}h, \qquad \varepsilon^* = \kappa^{\frac{1}{2}}(\theta_3)^{-1}\varepsilon, \qquad R_1 = 2\kappa^{\frac{1}{2}}\theta_1 R a^{-2}. \tag{3.3}$$

Substituting Eq. (3.2) into Eq. (2.12)$_1$ yields (see Appendix A for details)

$$g_0 = -\frac{2^v R_v}{\Gamma(1+v)}\left\{\varepsilon^* + \mathrm{i}\left[\frac{1}{R_1}\frac{1+v-4\rho^2}{(1+v)(2+v)} - h^*\right]\right\}, \tag{3.4a}$$

$$g_1 = -\frac{2^{1+v}[(1+v)^2 + 4\rho^2]\rho R_v}{R_1 \Gamma(4+v)}, \tag{3.4b}$$

$$g_2 = -\frac{\mathrm{i}}{R_1}\frac{R_v[(1+v)^2 + 4\rho^2][(3+v)^2 + 4\rho^2]}{2^{2-v}\Gamma(5+v)}, \qquad g_n = 0, \quad (n \geq 3) \tag{3.4c}$$

where

$$R_v = \Gamma(1 - w + \mathrm{i}\rho)\Gamma(1 - w - \mathrm{i}\rho) \tag{3.5}$$

is real[2].

Letting

$$J_k = \int_{-1}^{1} \eta^k \chi(\eta) \mathrm{d}\eta, \qquad k = 0,1 \tag{3.6}$$

the boundary conditions

$$\int_{-a}^{a} p(x)\mathrm{d}x = F\sin\varphi = P, \qquad \int_{-a}^{a} q(x)\mathrm{d}x = F\cos\varphi = Q \tag{3.7}$$

can be rewritten in the following compact form:

---

[2]This is because $\overline{\Gamma(z)} = \Gamma(\bar{z})$, where $z$ is a complex number.

$$\kappa^{\frac{1}{2}}P + i\kappa^{-\frac{1}{2}}Q = a^\nu J_0, \tag{3.8}$$

where the explicit expression of $J_0$ is given in Eq. (A6). Then it follows that

$$g_0 = \sigma_\nu a^{-\nu}\left(i\kappa^{\frac{1}{2}}P - \kappa^{-\frac{1}{2}}Q\right). \tag{3.9}$$

Taking Eqs. (2.10)$_2$, (2.12)$_2$, (3.4b, c) and (3.9) into account, we have

$$\frac{g_0 P_0^\rho(\xi)}{\lambda_0} = \frac{\left(i\kappa^{\frac{1}{2}}P - \kappa^{-\frac{1}{2}}Q\right)\sigma_\nu}{(2a)^\nu R_\nu}\Gamma(1+\nu), \tag{3.10a}$$

$$\frac{g_1 P_1^\rho(\xi)}{\lambda_1} = -\frac{4\rho[\xi(1+\nu) - 2i\rho]}{(3+\nu)(1+\nu)\nu R_1}, \tag{3.10b}$$

$$\frac{g_2 P_2^\rho(\xi)}{\lambda_2} = -\frac{16i}{R_1}\frac{1}{(3+\nu)(2+\nu)(1+\nu)\nu}P_2^\rho(\xi), \tag{3.10c}$$

$$\frac{g_n P_n^\rho(\xi)}{\lambda_n} = 0, \qquad (n \geq 3) \tag{3.10d}$$

where

$$P_2^\rho(\xi) = -\frac{3 + \nu + 4\rho^2}{8} + \xi^2\frac{(2+\nu)(3+\nu)}{8} - i\rho\xi\left(1+\frac{\nu}{2}\right). \tag{3.11}$$

Then by substituting Eq. (3.10) into (2.9) and recalling Eqs. (2.8a, b) and (3.3)$_3$, we can obtain the following expressions of the interfacial normal and tangential tractions inside the contact region

$$p(x) = \frac{1}{(a^2 - x^2)^{\frac{1-\nu}{2}}}\left\{I_1 \cos\left(\rho\ln\frac{a+x}{a-x}\right) - I_2 \sin\left(\rho\ln\frac{a+x}{a-x}\right)\right\}, \tag{3.12a}$$

$$q(x) = \frac{\kappa}{(a^2 - x^2)^{\frac{1-\nu}{2}}}\left\{I_2 \cos\left(\rho\ln\frac{a+x}{a-x}\right) + I_1 \sin\left(\rho\ln\frac{a+x}{a-x}\right)\right\}, \tag{3.12b}$$

where

$$I_1 = \frac{\Gamma(1+\nu)}{(2a)^\nu R_\nu}P + \frac{1}{\kappa\theta_1 R}\frac{a^2(1+4\rho^2) - x^2(2+\nu)}{(2+\nu)(1+\nu)\nu\sigma_\nu}, \tag{3.13a}$$

$$I_2 = \frac{\Gamma(1+\nu)}{(2a)^\nu R_\nu \kappa}Q + \frac{1}{\kappa\theta_1 R}\frac{2\rho a x}{(1+\nu)\nu\sigma_\nu}. \tag{3.13b}$$

With use of the identities listed in Eq. (A8), it can be verified that the interfacial tractions described in Eq. (3.12) satisfy the boundary conditions defined in Eq. (3.7). Oscillating with increasing frequency as $x$ approaches the contact edge, the tractions have the singularities of $r^{(v-1)/2+i\rho}$ at the contact perimeter $x = \pm a$. This immediately reminds us the similar stress singularity occurring at the interfacial cracks between two dissimilar materials. Compared with the case for homogenous materials ($v = 0$), the singularity of the stress has been weakened for graded materials (Giannakopoulos and Pallot, 2000).

In the homogeneous limit, i.e. $v \to 0$, one can show that

$$\beta \to 1, \quad \lambda \to 0, \quad w \to \frac{1}{2}, \quad \kappa \to 1, \quad \rho \to \frac{1}{2\pi}\ln(3-4\mu), \quad C_v \to \frac{2}{\pi}$$

$$R_v \to \frac{\pi\sqrt{3-4\mu}}{2(1-\mu)}, \quad \theta_1 \to \frac{(1+\mu)(1-2\mu)}{2E_0}, \quad v\sigma_v \to \frac{2\sqrt{3-4\mu}}{1-2\mu}. \tag{3.14}$$

Hence Eq. (3.12) reduces to

$$p(x) = \frac{1}{(a^2-x^2)^{\frac{1}{2}}}\left\{I_1' \cos\left(\rho_0 \ln\frac{a+x}{a-x}\right) - I_2' \sin\left(\rho_0 \ln\frac{a+x}{a-x}\right)\right\}, \tag{3.15a}$$

$$q(x) = \frac{1}{(a^2-x^2)^{\frac{1}{2}}}\left\{I_2' \cos\left(\rho_0 \ln\frac{a+x}{a-x}\right) + I_1' \sin\left(\rho_0 \ln\frac{a+x}{a-x}\right)\right\}, \tag{3.15b}$$

respectively. In Eq. (3.15)

$$I_1' = \frac{2P(1-\mu)}{\pi\sqrt{3-4\mu}} + \frac{a^2\left(\frac{1}{2}+2\rho_0^2\right)-x^2}{R(1+\mu)\sqrt{3-4\mu}}E_0, \tag{3.16a}$$

$$I_2' = \frac{2Q(1-\mu)}{\pi\sqrt{3-4\mu}} + \frac{2E_0 a\rho_0 x}{R(1+\mu)\sqrt{3-4\mu}} \tag{3.16b}$$

with

$$\rho_0 = \lim_{v \to 0} \rho = \frac{1}{2\pi}\ln(3-4\mu). \tag{3.16c}$$

In Appendix B, it can be shown that the interfacial tractions in Eq. (3.15) are exactly the same as that obtained by solving the corresponding Riemann-Hilbert problem using analytical function theory for homogeneous half-space.

With use of above results, the relations between the interfacial displacements and the corresponding external force can be established as follows

$$h = \frac{\theta_1 \sigma_v \kappa \Gamma(1+v)}{(2a)^v R_v} P + \frac{a^2}{2R} \frac{1+v-4\rho^2}{(1+v)(2+v)}, \quad \varepsilon = \frac{\theta_1 \sigma_v \Gamma(1+v)}{(2a)^v \kappa R_v} Q. \tag{3.17}$$

The elastic strain energy stored in the substrate is

$$U_E = U_{Eq} + U_{Ep}, \tag{3.18a}$$

where

$$U_{Eq} = \frac{1}{2} \int_{-a}^{a} q(x) \bar{u}_x \mathrm{d}x = \frac{1}{2} \int_{-a}^{a} q(x) \varepsilon \mathrm{d}x = \frac{1}{2} Q\varepsilon, \tag{3.18b}$$

$$U_{Ep} = \frac{1}{2} \int_{-a}^{a} p(x) \bar{u}_y \mathrm{d}x = \frac{1}{2} Ph - \frac{1}{4R} \int_{-a}^{a} x^2 p(x) \mathrm{d}x. \tag{3.18c}$$

Making use of the identities in Eq. (A8) and inserting Eq. (3.15a) into the last integral on the right-hand side of Eq. (3.18c), we have

$$\int_{-a}^{a} x^2 p(x) \mathrm{d}x = d_1 P a^2 - \frac{2^{1+v} R_v d_2}{R\kappa \theta_1 (4+v) v \sigma_v} a^{4+v}, \tag{3.19}$$

where

$$d_1 = \frac{1+v-4\rho^2}{(1+v)(2+v)}, \quad d_2 = \frac{(1+4\rho^2)[(1+v)^2 + 4\rho^2]}{(1+v)(2+v)\Gamma(3+v)} \tag{3.20}$$

are two dimensionless parameters.

The balance between elastic strain energy and surface energy yields that

$$\left.\frac{\partial(U_{Ep} + U_{Eq})}{\partial a}\right|_{\varepsilon,h} = 2\Delta\gamma. \tag{3.21}$$

Since

$$\left.\frac{\partial U_{Eq}}{\partial a}\right|_{\varepsilon} = \frac{1}{2}\left.\frac{\partial Q}{\partial a}\right|_{\varepsilon} \varepsilon, \tag{3.22a}$$

$$\left.\frac{\partial U_{Ep}}{\partial a}\right|_h = \frac{1}{2}\left.\frac{\partial P}{\partial a}\right|_h h - \frac{d_1}{4R}\left(a^2 \left.\frac{\partial P}{\partial a}\right|_h + 2Pa\right) + \frac{1}{4R^2}\frac{2^{1+v}R_v}{\kappa\theta_1 v\sigma_v}d_2 a^{3+v}, \qquad (3.22b)$$

then from Eq. (3.17), we have

$$\left.\frac{\partial P}{\partial a}\right|_h = \frac{v}{a}P - \frac{2^v R_v}{\theta_1\sigma_v\kappa R}\frac{1+v-4\rho^2}{\Gamma(3+v)}a^{1+v}, \qquad \left.\frac{\partial Q}{\partial a}\right|_\varepsilon = \frac{v}{a}Q. \qquad (3.23)$$

Substitution of Eqs. (3.7), (3.17), (3.22) and (3.23) into Eq. (3.21) leads to

$$\frac{d_3 v\sigma_v \Gamma(1+v)}{2^{1+v}R_v}(\kappa\sin^2\varphi + \kappa^{-1}\cos^2\varphi)\alpha^{-v}\frac{\Delta\gamma}{E^*R}\left(\frac{F}{\Delta\gamma}\right)^2\left(\frac{a}{R}\right)^{-(1+v)} - d_1\frac{a}{R}\frac{F}{\Delta\gamma}\sin\varphi$$

$$+ \frac{2^{v-1}d_2 R_v}{d_3 v\sigma_v \kappa}\alpha^v\frac{E^*R}{\Delta\gamma}\left(\frac{a}{R}\right)^{3+v} - 2 = 0, \qquad (3.24)$$

where

$$d_3 = -\frac{C_v}{v}\cos\frac{\beta\pi}{2}, \qquad E^* = \frac{E_0}{1-\mu^2}, \qquad \alpha = \frac{R}{c_0}. \qquad (3.25)$$

Eq. (3.24) describes the relationship between the normalized load $F/\Delta\gamma$ and the normalized contact half width $a/R$. For homogeneous half-space, Eq. (3.24) reduces to

$$\frac{1}{\pi}\frac{\Delta\gamma}{E^*R}\left(\frac{F}{\Delta\gamma}\right)^2\left(\frac{a}{R}\right)^{-1} - \frac{1-4\rho_0^2}{2}\frac{a}{R}\frac{F}{\Delta\gamma}\sin\varphi + \frac{\pi(1+4\rho_0^2)^2}{16}\frac{E^*R}{\Delta\gamma}\left(\frac{a}{R}\right)^3 - 2 = 0, \qquad (3.26)$$

by letting $v \to 0$.

When $\varphi = \pi/2$, the vertical force $P$ applied on the cylinder can be expressed as

$$P = \frac{(1-4\rho_0^2)\pi E^* a^2}{4R} - \sqrt{2\pi E^* a \Delta\gamma - \left(\frac{\rho_0 \pi E^* a^2}{R}\right)^2}, \qquad (3.27)$$

where $\rho_0 = (1/2\pi)\ln(3-4\mu)$. As shown in Eq. (3.27), which is derived from the non-slipping model, the effect of tangential traction manifests itself just through the term of $\rho_0$. For an incompressible half-space ($\mu = 0.5$), $\rho_0 = 0$ and Eq. (3.27) reduces to the following two-dimensional JKR-like solution

$$P = \pi E^* a^2/(4R) - \sqrt{2\pi E^* a \Delta\gamma}, \qquad (3.28a)$$

which was obtained by Barquins (1988) in the frictionless contact condition. This is consistent with the conclusion documented in Johnson (1985). From Eq. (3.28a), it can be obtained that (Barquins, 1988)

$$a_{\mathrm{JKR}}^{\mathrm{2D}} = \left(\frac{2\Delta\gamma R^2}{\pi E^*}\right)^{1/3} \tag{3.28b}$$

and

$$P_{\mathrm{JKR}}^{\mathrm{2D}} = 3\left(\frac{\pi E^* R \Delta\gamma^2}{16}\right)^{1/3}. \tag{3.28c}$$

## 3.2 Contact behavior with bending effect taken into account

In this subsection, adhesive contact behavior of graded half-space with bending moment taken into account will be discussed. This is a more realistic case since bending moment always exists unless the action point of the external force is properly chosen (Yao et al., 2008).

When bending moment is considered, it is necessary to include the displacement components arising from the rotation of cylinder into $\bar{u}_x$ and $\bar{u}_y$. Denoting the rigid body rotation angle of the cylinder as $\theta$, then under small deformation assumption, the interfacial displacements can be written as

$$\bar{u}_x = \varepsilon, \quad \bar{u}_y = h + \theta x - \frac{x^2}{2R}, \qquad |x| \leq a, \tag{4.1}$$

Replacing $x$ with $a\xi$ in Eq. (4.1) and inserting it to Eq. (2.8d) yield to

$$g(\xi) = -\frac{\mathrm{i}}{R_1}\xi^2 + \mathrm{i}\theta^*\xi + (\mathrm{i}h^* - \varepsilon^*), \tag{4.2}$$

where $h^*$, $\varepsilon^*$ and $R_1$ are defined in Eq. (3.3) and

$$\theta^* = \kappa^{-\frac{1}{2}}(\theta_1)^{-1} a \theta. \tag{4.3}$$

Substituting Eq. (4.2) into Eq. (2.12)$_1$ yields (see Appendix A for details)

$$g_0 = -\frac{2^v R_v}{\Gamma(1+v)}\left\{\varepsilon^* - \frac{2\rho\theta^*}{1+v} + i\left[\frac{1}{R_1}\frac{1+v-4\rho^2}{(1+v)(2+v)} - h^*\right]\right\}, \quad (4.4a)$$

$$g_1 = \frac{[(1+v)^2 + 4\rho^2]R_v}{2^{1-v}\Gamma(3+v)}\left(i\theta^* - \frac{4\rho}{R_1}\frac{1}{3+v}\right), \quad (4.4b)$$

$$g_2 = -\frac{i}{R_1}\frac{R_v[(1+v)^2+4\rho^2][(3+v)^2+4\rho^2]}{2^{2-v}\Gamma(5+v)}, \quad g_n = 0 \quad (n \geq 3). \quad (4.4c)$$

Recalling Eq. (3.6), the boundary conditions

$$\int_{-a}^{a} p(x)\mathrm{d}x = P, \quad \int_{-a}^{a} q(x)\mathrm{d}x = Q, \quad \int_{-a}^{a} xp(x)\mathrm{d}x = M \quad (4.5)$$

can be rewritten as

$$\kappa^{\frac{1}{2}}P + i\kappa^{-\frac{1}{2}}Q = a^v J_0, \quad \kappa^{\frac{1}{2}}M = a^{1+v}\mathrm{Re}(J_1), \quad (4.6)$$

where the explicit expressions of $J_0$ and $J_1$ are given in Eqs. (A6) and (A7), respectively.

Substituting Eqs. (A6), (A7), (2.12)$_2$ and (4.4a, b) into Eq. (4.6) yields

$$\kappa^{\frac{1}{2}}P + i\kappa^{-\frac{1}{2}}Q = \frac{(2a)^v R_v}{\Gamma(v+1)\sigma_v}\left[i\left(\varepsilon^* - \frac{2\rho\theta^*}{1+v}\right) - \left(\frac{d_1}{R_1} - h^*\right)\right], \quad (4.7a)$$

$$\frac{\kappa^{\frac{1}{2}}\Gamma(2+v)M}{(2a)^{v+1}R_v} = \frac{(1+v)^2 + 4\rho^2}{2(2+v)(1+v)v\sigma_v}\theta^* + \frac{\rho}{\sigma_v}\left(-\varepsilon^* + \frac{2\rho\theta^*}{1+v}\right), \quad (4.7b)$$

where $d_1$ is defined in Eq. (3.20)$_1$.

Then, combining Eqs. (3.3), (4.3) and (4.7) gives rise to

$$h = \frac{\theta_1\sigma_v\kappa\Gamma(1+v)}{(2a)^v R_v}P + d_1\frac{a^2}{2R}, \quad (4.8a)$$

$$\varepsilon = \frac{\sigma_v\theta_1\Gamma(3+v)}{[(1+v)^2 + 4\rho^2]R_v}\left[\frac{(1+4\rho^2)(1+v)}{2+v}\frac{Q}{(2a)^v} + \frac{4\rho v M}{(2a)^{1+v}}\right], \quad (4.8b)$$

$$\theta = \frac{4\theta_1 v\sigma_v\Gamma(3+v)}{[(1+v)^2+4\rho^2]R_v}\left[\frac{\rho Q}{(2a)^{1+v}} + \frac{(1+v)\kappa M}{(2a)^{2+v}}\right]. \quad (4.8c)$$

From Eq. (4.8), it can be seen that the bending moment $M$ has no influence on the maximum indentation depth $h$, but will affect both the tangential displacement $\varepsilon$ and the angular rotation $\theta$.

Furthermore, by combining Eqs. (2.4), (2.8a, b), (2.12)$_2$, (3.4b), (4.4) and (4.8), we have

$$p(x) = \frac{1}{(a^2 - x^2)^{\frac{1-v}{2}}}\left\{I_3 \cos\left(\rho\ln\frac{a+x}{a-x}\right) - I_4 \sin\left(\rho\ln\frac{a+x}{a-x}\right)\right\}, \quad (4.9a)$$

$$q(x) = \frac{\kappa}{(a^2 - x^2)^{\frac{1-v}{2}}}\left\{I_4 \cos\left(\rho\ln\frac{a+x}{a-x}\right) + I_3 \sin\left(\rho\ln\frac{a+x}{a-x}\right)\right\}, \quad (4.9b)$$

where

$$I_3 = \frac{\Gamma(v+1)}{(2a)^v R_v}P$$
$$+ \frac{2x\Gamma(3+v)}{a[(1+x)^2 + 4\rho^2]R_v}\left[\frac{(1+v)M}{(2a)^{1+v}} + \frac{\rho Q}{\kappa(2a)^v}\right] + \frac{1}{\sigma_v R\kappa\theta_1}\left[\frac{1+4\rho^2}{(2+v)(1+v)v}a^2 - \frac{x^2}{v(1+v)}\right],$$

$$I_4 = \frac{\Gamma(v+1)}{(2a)^v R_v}\frac{Q}{\kappa} - \frac{4\rho\Gamma(3+v)}{[(1+x)^2 + 4\rho^2]R_v}\left[\frac{M}{(2a)^{1+v}} + \frac{1}{1+v}\frac{\rho Q}{\kappa(2a)^v}\right] + \frac{1}{\sigma_v R\kappa\theta_1}\frac{2\rho ax}{v(1+v)}.$$

(4.10a, b)

Following the same procedures described in the previous section, the governing equation for the dimensionless contact half width can be obtained as

$$(A_1\sin^2\varphi + A_5\cos^2\varphi)\alpha^{-v}\frac{\Delta\gamma}{E^*R}\left(\frac{F}{\Delta\gamma}\right)^2\left(\frac{a}{R}\right)^{-(1+v)} - A_2\sin\varphi\frac{F}{\Delta\gamma}\frac{a}{R}$$

$$+ A_3\alpha^{-v}\frac{\Delta\gamma}{E^*R}\left(\frac{M}{R\Delta\gamma}\right)^2\left(\frac{a}{R}\right)^{-(3+v)} + A_4\cos\varphi\alpha^{-v}\frac{\Delta\gamma}{E^*R}\frac{M}{R\Delta\gamma}\frac{F}{\Delta\gamma}\left(\frac{a}{R}\right)^{-(2+v)}$$

$$+ A_6\alpha^v\frac{E^*R}{\Delta\gamma}\left(\frac{a}{R}\right)^{3+v} - 4 = \mathcal{R}\left(\frac{F}{\Delta\gamma}, \frac{M}{R\Delta\gamma}, \frac{a}{R}\right) = 0, \quad (4.11)$$

where the coefficients $A_i$, $i = 1,2,\ldots,6$ defined as

$$A_1 = \frac{\sigma_v d_3}{2^v R_v}\kappa v\Gamma(1+v), \qquad A_2 = 2d_1, \quad (4.12a)$$

$$A_3 = \frac{\sigma_v d_3}{2^v R_v}\frac{v(1+v)(2+v)\Gamma(3+v)\kappa}{(1+v)^2 + 4\rho^2}, \qquad A_4 = \frac{\sigma_v d_3}{2^v R_v}\frac{4\rho v(1+v)\Gamma(3+v)}{(1+v)^2 + 4\rho^2}, \quad (4.12b)$$

$$A_5 = \frac{\sigma_v d_3}{2^v R_v}\frac{v(1+v)(1+4\rho^2)\Gamma(2+v)}{[(1+v)^2 + 4\rho^2]\kappa}, \qquad A_6 = \frac{2^v(4+v)R_v d_4}{v\sigma_v d_3\kappa} \quad (4.12c)$$

are all dimensionless quantities. $d_1$, $d_3$ and $\alpha$ in Eq. (4.12) are defined in Eqs. (3.20)$_1$,

$(3.25)_1$ and $(3.25)_3$, respectively. Meanwhile,

$$d_4 = \frac{(3+v)(1+4\rho^2)[(1+v)^2+4\rho^2]}{(1+v)(2+v)\Gamma(5+v)}. \tag{4.13}$$

Eq. (4.11) describes an equilibrium surface $\mathcal{R}\left(\frac{F}{\Delta\gamma},\frac{M}{R\Delta\gamma},\frac{a}{R}\right)=0$ in $a/R - F/\Delta\gamma - M/R\Delta\gamma$ space implicitly. The "edge" curve of this equilibrium surface, which represents the critical state, can be determined from the following formula:

$$\begin{cases} \mathcal{R}\left(\frac{F}{\Delta\gamma},\frac{M}{R\Delta\gamma},\frac{a}{R}\right)=0 \\ \partial\mathcal{R}\left(\frac{F}{\Delta\gamma},\frac{M}{R\Delta\gamma},\frac{a}{R}\right)/\partial a=0 \end{cases}. \tag{4.14}$$

Generally, a specific loading condition can be represented by an implicit function $\mathcal{B}(F/\Delta\gamma,M/R\Delta\gamma)=0$. Then the corresponding critical state (if it exists) can be determined by solving (4.14) and $\mathcal{B}(F/\Delta\gamma,M/R\Delta\gamma)=0$ simultaneously. See Fig. 2 for reference.

For a homogeneous half-space, Eq. (4.11) reduces to

$$\frac{1}{\pi}\frac{\Delta\gamma}{E^*R}\left(\frac{F}{\Delta\gamma}\right)^2\left(\frac{a}{R}\right)^{-1} - \frac{1-4\rho_0^2}{2}\frac{a}{R}\frac{F}{\Delta\gamma}\sin\varphi + \frac{\pi(1+4\rho_0^2)^2}{16}\frac{E^*R}{\Delta\gamma}\left(\frac{a}{R}\right)^3$$
$$+\frac{4}{\pi(1+4\rho_0^2)}\frac{\Delta\gamma}{E^*R}\left(\frac{M}{R\Delta\gamma}\right)^2\left(\frac{a}{R}\right)^{-3} + \frac{8\rho_0\cos\varphi}{\pi(1+4\rho_0^2)}\frac{\Delta\gamma}{E^*R}\frac{M}{R\Delta\gamma}\frac{F}{\Delta\gamma}\left(\frac{a}{R}\right)^{-2} - 2 = 0.$$
$$\tag{4.15}$$

Comparing Eq. (4.15) with Eq. (3.26), it seems that the bending moment $M$ affects the contact width $a$ only through the fourth and fifth terms. When $M=0$, Eq. (4.15) will reduce to Eq. (3.26). On the other hand, if $F=0$ (there is no net force exerted on the cylinder), from Eq. (4.11), we have

$$A_3\alpha^{-v}\frac{\Delta\gamma}{E^*R}\left(\frac{M}{R\Delta\gamma}\right)^2\left(\frac{a}{R}\right)^{-(3+v)} + A_6\alpha^v\frac{E^*R}{\Delta\gamma}\left(\frac{a}{R}\right)^{3+v} - 4 = 0, \tag{4.16}$$

which can be simplified as

$$\frac{|M|}{R\Delta\gamma} = \sqrt{\frac{4\left(\frac{a}{R}\right)^{3+v} - K_1\left(\frac{a}{R}\right)^{6+2v}}{K_2}}, \quad (4.17)$$

where

$$K_1 = A_6 \alpha^v \frac{E^*R}{\Delta\gamma}, \qquad K_2 = A_3 \alpha^{-v} \frac{\Delta\gamma}{E^*R}. \quad (4.18)$$

The critical contact half width at bend-off can be obtained from the following condition

$$\frac{\partial M}{\partial a} = 0, \quad (4.19)$$

which gives

$$a_{\text{bend-off}} = R\left(\frac{2}{K_1}\right)^{\frac{1}{3+v}}. \quad (4.20)$$

Inserting Eq. (4.20) back into Eq. (4.17) yields the closed-form solution for the bend-off moment which is required to bend the cylinder away from the substrate

$$M_{\text{bend-off}} = \pm\frac{2R\Delta\gamma}{\sqrt{1+4\rho^2}}. \quad (4.21)$$

In the homogeneous limit, the critical contact half width and the corresponding bend-off moment reduce to

$$a_{\text{bend-off}}^{\text{homo}} = \left(\frac{16}{\pi(1+4\rho_0^2)^2}\frac{R^2\Delta\gamma}{E^*}\right)^{\frac{1}{3}}, \qquad M_{\text{bend-of}}^{\text{homo}} = \pm\frac{2R\Delta\gamma}{\sqrt{1+4\rho_0^2}}, \quad (4.22)$$

respectively.

From the above analysis, it is interesting to note that the bend-off moment is independent of the two parameters $E^*R/\Delta\gamma$ and $\alpha$ defined in Eq. (3.25)$_3$. In other words, it is irrespective of the characteristic depth $c_0$ and the characteristic Young's modulus $E_0$ but only depends on the gradient exponent $v$, Poisson ratio $\mu$, cylinder radius $R$ and the work of adhesion $\Delta\gamma$.

At this position, it is worth noting that for any symmetric indenter whose shape can be described by a polynomial, the closed-form analytical solution of the corresponding problem can also be obtained with the above procedure. This is because it can be proved, with use of the properties of Jacobi polynomials, that there are only finite non-zero terms of $g_m$ in Eq. (2.9) if $g(\xi)$ is a polynomial function of $\xi$.

## 4. Results and discussions

First, let us investigate the adhesive contact behavior under vertical force (i.e., $\varphi = \pi/2$). For comparison purpose, the variation of the contact width under non-slipping condition with coupling effect taken into account along with the corresponding solutions for frictionless contact (Chen et al., 2009a) are schematically shown in Fig. 3a and Fig. 3b. It can be observed that the two sets of solutions agree well in the tensile regime ($P < 0$). Whereas, in the compressive regime ($P > 0$), they give distinct predictions of the contact width. Concretely speaking, it is found that the tangential tractions due to friction can reduce the area of contact between the cylinder and substrate. Compared with graded materials, this difference is more obvious for homogenous case. This is reasonable since compared with homogenous material, the average stiffness near the surface of the half-space is larger for graded materials whose modulus are varied in the form of $E = E_0(z/c_0)^v$, $0 < v < 1$ along depth. Figure 3 also shows that the pull-off forces are almost identical for homogenous material and the graded materials tested.

Similar conclusions can also be drawn from Fig. 4, which depicts the normalized inclined force as a function of the normalized contact half width for different pulling angles under a specific set of parameters $v = 0.1$, $\mu = 0.3, \alpha = 100$, and $E^*R/\Delta\gamma = 100$. This figure shows that the coupling effect between the normal and tangential problem becomes obvious when the external force is inside the compressive regime ($F > 0$). In addition, the critical force corresponding to $\varphi = 0$ (only shear force applied) exhibits bilateral symmetry about contact center.

The above comparison results indicate that the coupling effect almost does not show significant influence on the critical force and critical contact width at pull-off. Therefore when the bending effect is neglected, nearly the same results as those of Chen et al. (2009a) are obtained at pull-off. Figure 5 plots the pull-off force (scaled by $P_{\text{JKR}}^{\text{2D}}$ defined in Eq. (3.28c)) and the critical contact width (scaled by $a_{\text{JKR}}^{\text{2D}}$ defined in Eq. (3.28b)) as functions of the gradient exponent $v$ for various α with the other parameters are constant. Figure 6 shows the effect of $E^*R/\Delta\gamma$ on the critical force and the critical contact width. From this figure, it is observed that parameter $E^*R/\Delta\gamma$ does not significantly influence the trend of the variations of critical contact width or pull-off force with respect to $v$ even though coupling effect is considered. Figure 7 describes the variation of the normalized pull-off force $F_{\text{pull-off}}/P_{\text{JKR}}^{\text{2D}}$ as a function of the pulling angle $\varphi$ for materials with different values of $v$ and $\alpha$. We refer the readers to Chen et al. (2009a) for more detail discussions.

When bending moment is considered, the corresponding adhesive behavior is shown in Fig. 8. Viewed from different angles in $a/R - F/\Delta\gamma - M/R\Delta\gamma$ space, Figures 8(a, b) plot an equilibrium surface with a specific set of parameters described by Eq. (4.11). Figure 8c shows the projection of the curves obtained by intersecting the surface with a $M/R\Delta\gamma = \text{Const}$ plane onto the $a/R - F/\Delta\gamma$ plane. From these figures, it is observed that bending moment has a significant effect on the contact behavior of the considered system. For a specific loading process, it can be represented by a curve on the equilibrium surface. When this curve arrived at the edge of the surface, the cylinder will be taken away from the substrate.

For the case where only bending moment exists ($F = 0$), Figure 9 plots the relation between the contact width and the bending moment for prescribed values of $E^*R/\Delta\gamma$ and $\mu$. It shows that the contact width is independent of the direction of the bending moment. In addition, one can also notice that each curve has almost the same critical point ($M/(R\Delta\gamma) \approx \pm 2$), at which the contact is separated. Given bend-off moment, adhesion strength is measured by the critical contact area. Therefore, it can be seen from Fig. 9 that for a specific gradient exponent, the larger the value of $\alpha$, the more difficult to bend the cylinder apart from the graded substrate.

Figure 10 plots the variation of the normalized bend-off moment $M_{\text{bend-off}}/M_{\text{bend-off}}^{\text{homo}}$ with respect to the gradient exponent $v$ for prescribed Young's modulus $\mu = 0.3$. It shows that the bend-off moment changes slightly although it rises monotonically with the increasing value of $v$. It means that the bend-off moment is insensitive to the gradient exponent $v$.

Figure 11 shows the influence of $\alpha$ and $v$ on the critical contact width at bend-off when the values of $E^*R/\Delta\gamma$ and $\mu$ are prescribed. It is interesting that the trend of critical contact width at bend-off shares the similar character with that corresponding to the pull-off force plotted in Fig. 5b.

The effect of $E^*R/\varDelta\gamma$ on the critical contact width at bend-off is shown in Fig. 12. One can see that parameter $E^*R/\varDelta\gamma$ does not significantly affect the critical contact width, just as we concluded from Fig. 6b.

## 5. Summary

In this paper, we examined the non-slipping adhesive contact behavior of power-law graded materials focusing on the coupling effect between the tangential and normal directions which was often neglected in previous works. Based on the theory developed by Popov (1972), some closed-form analytical solutions were obtained for the considered problem. Our analysis showed that the coupling effect becomes obvious mainly in the compressive regime of the external force but does not show great influence on the critical force and critical contact width at pull-off. This justified some of the results obtained by Chen et al. (2009a) about the adhesive behavior of graded materials where the coupling effect was ignored.

The role of the bending moment was also studied analytically in the present paper. The closed-form analytical solutions of the critical contact width and critical moments at bending off are obtained. Our analysis showed that bending moment has a significant effect on the adhesive behavior of the considered system. When only bending moment is applied, unlike pull-off force, the bend-off moment is not sensitive to the gradient exponent $v$ and is almost equal to that for the corresponding homogeneous case. On the other hand, the trend of critical contact width at bend-off shares the similar character with that at pull-off. Compared to the homogeneous case, the graded materials has critical contact width which is larger for small values of $\alpha$ but smaller for large values of $\alpha$. The results obtained in this paper are helpful for understanding the adhesive behavior of contact systems involving graded elastic materials.

**Acknowledgements**

The financial supports from the National Natural Science Foundation (10925209, 10772037, 10472022, 10721062) and the National Key Basic Research Special Foundation of China (2006CB601205) are gratefully acknowledged.

**Appendix A: Some Integrals related to Jacobi Polynomials**

In this appendix, some useful integrals of Jacobi polynomials (e.g. $g_m$ in Eq. (2.12)$_1$ and $J_k$ in Eq. (3.6)) are derived.

First, let us introduce the following two identities

$$\int_{-1}^{1} (1-x)^\rho (1+x)^\beta P_m^{(\alpha,\beta)}(x) = \frac{2^{\beta+\rho+1}\Gamma(\rho+1)\Gamma(\beta+m+1)\Gamma(\alpha-\rho+m)}{m!\,\Gamma(\alpha-\rho)\Gamma(\beta+\rho+m+2)},$$

$$\text{for } \operatorname{Re}\rho > -1, \operatorname{Re}\beta > -1. \quad \text{(A1)}$$

$$\int_{-1}^{1} (1-x)^\rho (1+x)^\sigma P_m^{(\alpha,\beta)}(x) = \frac{2^{\beta+\sigma+1}\Gamma(\rho+1)\Gamma(\sigma+1)\Gamma(m+1+\alpha)}{m!\,\Gamma(\rho+\sigma+2)\Gamma(1+\alpha)}$$

$$\times\; _3F_2(-m, \alpha+\beta+m+1, \rho+1; \alpha+1, \rho+\sigma+2; 1), \quad \text{for } \operatorname{Re}\rho > -1, \operatorname{Re}\beta > -1. \quad \text{(A2)}$$

where $_3F_2(a,b,c;d,e;z)$ is the hypergeometric function.

With use of the above results, it can be shown that

$$\int_{-1}^{1} \frac{P_0^{-\rho}(\xi)}{\psi_{-\rho}(\xi)} d\xi = \frac{2^v R_v}{\Gamma(1+v)}, \qquad \int_{-1}^{1} \frac{P_0^{-\rho}(\xi)}{\psi_{-\rho}(\xi)} \xi d\xi = -\frac{2^{1+v} i\rho R_v}{\Gamma(2+v)},$$

$$\int_{-1}^{1} \frac{P_0^{-\rho}(\xi)}{\psi_{-\rho}(\xi)} \xi^2 d\xi = \frac{2^v R_v[1+v-4\rho^2]}{\Gamma(3+v)},$$

$$\int_{-1}^{1} \frac{P_1^{-\rho}(\xi)}{\psi_{-\rho}(\xi)} d\xi = 0, \quad \int_{-1}^{1} \frac{P_1^{-\rho}(\xi)}{\psi_{-\rho}(\xi)} \xi\, d\xi = \frac{[(1+v)^2+4\rho^2]R_v}{2^{1-v}\Gamma(3+v)},$$

$$\int_{-1}^{1} \frac{P_1^{-\rho}(\xi)}{\psi_{-\rho}(\xi)} \xi^2 d\xi = -\frac{2^{1+v}i\rho[(1+v)^2+4\rho^2]R_v}{\Gamma(4+v)},$$

$$\int_{-1}^{1} \frac{P_2^{-\rho}(\xi)}{\psi_{-\rho}(\xi)} d\xi = \int_{-1}^{1} \frac{P_2^{-\rho}(\xi)}{\psi_{-\rho}(\xi)} \xi\, d\xi = 0,$$

$$\int_{-1}^{1} \frac{P_2^{-\rho}(\xi)}{\psi_{-\rho}(\xi)} \xi^2 d\xi = \frac{2^{v-2}R_v[(1+v)^2+4\rho^2][(3+v)^2+4\rho^2]}{\Gamma(5+v)},$$

$$\int_{-1}^{1} \frac{P_n^{-\rho}(\xi)}{\psi_{-\rho}(\xi)} d\xi = \int_{-1}^{1} \frac{P_n^{-\rho}(\xi)}{\psi_{-\rho}(\xi)} \xi\, d\xi = \int_{-1}^{1} \frac{P_n^{-\rho}(\xi)}{\psi_{-\rho}(\xi)} \xi^2\, d\xi = 0 \quad (n \geq 3) \quad \text{(A3)}$$

and

$$\int_{-1}^{1} \frac{P_0^\rho(\xi)}{\psi_\rho(\xi)} d\xi = \frac{2^v R_v}{\Gamma(1+v)}, \qquad \int_{-1}^{1} \frac{P_n^\rho(\xi)}{\psi_\rho(\xi)} d\xi = 0, \quad (n \geq 1)$$

$$\int_{-1}^{1} \frac{P_0^\rho(\xi)}{\psi_\rho(\xi)} \xi\, d\xi = \frac{2^{1+v} i\rho R_v}{\Gamma(2+v)}, \qquad \int_{-1}^{1} \frac{P_1^\rho(\xi)}{\psi_\rho(\xi)} \xi\, d\xi = \frac{[(1+v)^2 + 4\rho^2] R_v}{2^{1-v}\Gamma(3+v)},$$

$$\int_{-1}^{1} \frac{P_n^\rho(\xi)}{\psi_\rho(\xi)} \xi\, d\xi = 0, \quad (n \geq 2) \tag{A4}$$

where $P_m^\rho = P_m^{-w-i\rho,-w+i\rho}(\xi)$. $\psi_\rho$ and $R_v$ are defined in Eq. (2.10)$_1$ and Eq. (3.5), respectively. In the above derivations, the following facts

$$\frac{1}{\Gamma(0)} = 0, \qquad \Gamma(1+z) = z\Gamma(z). \tag{A5}$$

have been used. In (A5), $z$ is a complex number.

From the results in (A3), Eqs. (3.4) and (4.4) can be obtained by inserting Eqs. (3.2) and (4.2) into Eq. (2.12)$_1$, respectively. From the results in (A4), the integral in (3.6) can be calculated as:

$$J_0 = \int_{-1}^{1} \chi(\eta) d\eta = \sum_{m=0}^{\infty} \frac{g_m}{i\sigma_v \lambda_m} \int_{-1}^{1} \frac{P_m^\rho(\eta)}{\psi_\rho(\eta)} d\eta = \frac{g_0}{i\sigma_v \lambda_0} \frac{2^v R_v}{\Gamma(1+v)}, \tag{A6}$$

$$J_1 = \int_{-1}^{1} \chi(\eta)\eta\, d\eta = \sum_{m=0}^{\infty} \frac{g_m}{i\sigma_v \lambda_m} \int_{-1}^{1} \frac{P_m^\rho(\eta)}{\psi_\rho(\eta)} \eta\, d\eta$$

$$= \frac{2^{1+v} R_v}{i\sigma_v \Gamma(2+v)} \left[ \frac{(1+v)^2 + 4\rho^2}{4(2+v)} \frac{g_1}{\lambda_1} + i\rho \frac{g_0}{\lambda_0} \right]. \tag{A7}$$

Furthermore, by expressing the sine and cosine functions in complex exponential form and making use of Eq. (A1), we have

$$\int_{-a}^{a} \frac{x^{2n}}{(a^2 - x^2)^w} \sin\left(\rho \ln \frac{a+x}{a-x}\right) dx = 0, \qquad n = 0,1,2,\ldots$$

$$\int_{-a}^{a} \frac{x^{2n+1}}{(a^2 - x^2)^w} \cos\left(\rho \ln \frac{a+x}{a-x}\right) dx = 0, \qquad n = 0,1,2,\ldots$$

$$\int_{-a}^{a} \frac{1}{(a^2 - x^2)^w} \cos\left(\rho \ln \frac{a+x}{a-x}\right) dx = \frac{R_v}{\Gamma(1+v)} (2a)^v,$$

$$\int_{-a}^{a} \frac{x^2}{(a^2-x^2)^w} \cos\left(\rho \ln \frac{a+x}{a-x}\right) dx = \frac{1+v-4\rho^2}{\Gamma(3+v)} R_v 2^v a^{2+v},$$

$$\int_{-a}^{a} \frac{x^4}{(a^2-x^2)^w} \cos\left(\rho \ln \frac{a+x}{a-x}\right) dx = \frac{3(1+v)(3+v) - 8\rho^2(7+2v-2\rho^2)}{\Gamma(5+v)} R_v 2^v a^{4+v},$$

$$\int_{-a}^{a} \frac{x}{(a^2-x^2)^w} \sin\left(\rho \ln \frac{a+x}{a-x}\right) dx = \frac{2^{1+v} \rho R_v}{\Gamma(2+v)} a^{1+v},$$

$$\int_{-a}^{a} \frac{x^3}{(a^2-x^2)^w} \sin\left(\rho \ln \frac{a+x}{a-x}\right) dx = \frac{2\rho(5+3v-4\rho^2)}{\Gamma(4+v)} 2^v R_v a^{3+v}, \tag{A8}$$

where $w$ is defined in (2.10)$_2$.

**Appendix B: Tractions force inside the adhesive contact region of a rigid cylinder on homogeneous isotropic half-space**

In this appendix, the traction force distribution inside the adhesive contact region of a rigid cylinder on homogeneous isotropic half-space under vertical and horizontal force (denoted as $P$ and $Q$, respectively) will be derived for comparison purpose.

Since there is no mismatch strain in our problem, then by adopting the notations in Chen and Gao (2006), we have

$$(1-\beta)T_1^+ + (1+\beta)T_1^- = \frac{E^*}{4R}x, \tag{B1a}$$

$$(1+\beta)T_2^+ + (1-\beta)T_2^- = -\frac{E^*}{4R}x, \tag{B1b}$$

where

$$E^* = \frac{E}{1-\mu^2}, \quad \beta = \frac{1-2\mu}{2(1-\mu)} \tag{B2}$$

and $E$ and $\mu$ are the Young's modulus and Poisson's ratio of the half-space, respectively.

The Riemann-Hilbert equations in Eq. (B1) can be solved analytically by following the standard analytical function approach. By virtue of some calculations based on the identities given by Guo and Jin (2009), we have

$$T_1^+ - T_1^- = \frac{iE^*}{4R}\cosh(\pi\kappa)\left[x^2 - 2i\kappa ax - a^2\left(\frac{1}{2}+2\kappa^2\right)\right](a+x)^{-\bar{r}}(a-x)^{-r}$$
$$-2ic_1\cosh(\pi\kappa)(a+x)^{-\bar{r}}(a-x)^{-r}, \tag{B3a}$$

$$T_2^+ - T_2^- = -\frac{iE^*}{4R}\cosh(\pi\kappa)\left[x^2 + 2i\kappa ax - a^2\left(\frac{1}{2}+2\kappa^2\right)\right](a+x)^{-r}(a-x)^{-\bar{r}}$$
$$-2ic_2\cosh(\pi\kappa)(a+x)^{-r}(a-x)^{-\bar{r}}, \tag{B3b}$$

and

$$q(x) = (T_1^+ - T_1^-) + (T_2^+ - T_2^-), \quad p(x) = i(T_1^+ - T_1^-) - i(T_2^+ - T_2^-), \tag{B4}$$

where

$$r = \frac{1}{2} + i\kappa, \quad \kappa = \frac{1}{2\pi}\ln\frac{1+\beta}{1-\beta} \tag{B5}$$

and $q(x)$ and $p(x)$ denote the tangential and normal traction forces, respectively.

The constants $c_1$ and $c_2$ in Eq. (B3) can be determined by using the boundary conditions

$$\int_{-a}^{a} q(x)\mathrm{d}x = Q, \qquad \int_{-a}^{a} p(x)\mathrm{d}x = P, \tag{B6}$$

which gives

$$c_1 = \frac{P + iQ}{4\pi}, \qquad c_2 = -\frac{P - iQ}{4\pi}. \tag{B7}$$

Then by combining Eq. (B3), (B4) and (B7), we can obtain that

$$p(x) = \frac{1}{\sqrt{a^2 - x^2}}\left\{I_1'' \cos\left(\kappa\ln\frac{a+x}{a-x}\right) - I_2'' \sin\left(\kappa\ln\frac{a+x}{a-x}\right)\right\}, \tag{B8a}$$

$$q(x) = \frac{1}{\sqrt{a^2 - x^2}}\left\{I_2'' \cos\left(\kappa\ln\frac{a+x}{a-x}\right) + I_1'' \sin\left(\kappa\ln\frac{a+x}{a-x}\right)\right\}, \tag{B8b}$$

where

$$I_1'' = \frac{2P(1-\mu)}{\pi\sqrt{3-4\mu}} + \frac{a^2\left(\frac{1}{2} + 2\kappa^2\right) - x^2}{R(1+\mu)\sqrt{3-4\mu}} E, \tag{B9a}$$

$$I_2'' = \frac{2Q(1-\mu)}{\pi\sqrt{3-4\mu}} + \frac{2E a \kappa x}{R(1+\mu)\sqrt{3-4\mu}} \tag{B9b}$$

and

$$\kappa = \frac{1}{2\pi}\ln(3 - 4\mu). \tag{B10}$$

It is obvious the results in Eq. (B8) are the same as that in Eq. (3.15), which verifies the assertion made in Section 3 of the main text.

**Figure captions**

Figure 1

A rigid cylinder of radius $R$ in non-slipping contact with an elastic graded half-space with Young's modulus varying with depth $z$ according to $E = E_0(z/c_0)^v$ $(0 < v < 1)$ while Poisson's ratio remaining constant. Both external force $F$ and bending moment $M$ are applied.

Figure 2

Schematic illustration of the critical curve projected onto the $F/\Delta\gamma - M/R\Delta\gamma$ plane, loading path and the corresponding critical state.

Figure 3

Comparison of the contact half width as a function of the pulling force considering coupling effect with the corresponding results in the absence of tangible tractions for (a) different values of $\alpha$; (b) different values of $v$. Here $\varphi = \pi/2$, $\mu = 0.3$ and $E^*R/\Delta\gamma = 100$.

Figure 4

Comparison of the contact half width obtained by considering coupling effect with the corresponding results in the absence of coupling effect for different pulling angles. Here $v = 0.1$, $\mu = 0.3$, $E^*R/\Delta\gamma = 100$ and $\alpha = 100$.

Figure 5

The gradient exponent $v$ versus (a) the normalized pull-off force $P_{\text{pull-off}}/P_{\text{JKR}}^{\text{2D}}$ and (b) the normalized pull-off contact width $a_{\text{pull-off}}/a_{\text{JKR}}^{\text{2D}}$ with $\mu = 0.3$, $E^*R/\Delta\gamma = 100$ and different values of $\alpha$.

Figure 6

The gradient exponent $v$ versus (a) the normalized pull-off force $P_{\text{pull-off}}/P_{\text{JKR}}^{\text{2D}}$ and (b) the normalized pull-off contact width $a_{\text{pull-off}}/a_{\text{JKR}}^{\text{2D}}$ with $\mu = 0.3$ and different values of $E^*R/\Delta\gamma$ and $\alpha$.

Figure 7

Variation of the normalized pull-off force $F_{\text{pull-off}}/P_{\text{JKR}}^{\text{2D}}$ as a function of the pulling angle $\varphi$ for different values of gradient exponent $v$. (a) $\alpha = 1000$; (b) $\alpha = 100$; (c) $\alpha = 10$; (d) $\alpha = 2$. Here $\mu = 0.3$, $E^*R/\Delta\gamma = 100$.

Figure 8

The equilibrium surface described by Eq. (4.11) with a specific set of parameters in $a/R - F/\Delta\gamma - M/R\Delta\gamma$ space. (a) and (b) are side views from different angles. (c) Projections of the curves obtained by intersecting the surface with a $M/R\Delta\gamma = \text{Const}$ plane onto the $a/R - F/\Delta\gamma$ plane. Here $\mu = 0.3$, $E^*R/\Delta\gamma = 100$, $\alpha = 100$, $v = 0.1$ and $\varphi = \pi/2$.

Figure 9

The normalized external bending moment $M/(R\Delta\gamma)$ as a function of the normalized contact half width $a/R$ for homogeneous and graded materials ($v = 0.5$) with different values of $\alpha$. Here $\mu = 0.3$, $E^*R/\Delta\gamma = 100$.

Figure 10

The normalized bend-off moment $M_{\text{bend-off}}/M_{\text{bend-off}}^{\text{homo}}$ versus the gradient exponent $v$ with $\mu = 0.3$.

Figure 11

The normalized bend-off contact width $a_{\text{bend-off}}/a_{\text{bend-off}}^{\text{homo}}$ versus the gradient exponent $v$ with $\mu = 0.3$, $E^*R/\Delta\gamma = 100$ and different values of $\alpha$.

Figure 12

The normalized bend-off contact width $a_{\text{bend-off}}/a_{\text{bend-off}}^{\text{homo}}$ versus the gradient exponent $v$ under the effects of different values of $E^*R/\Delta\gamma$ and $\alpha$ with $\mu = 0.3$.

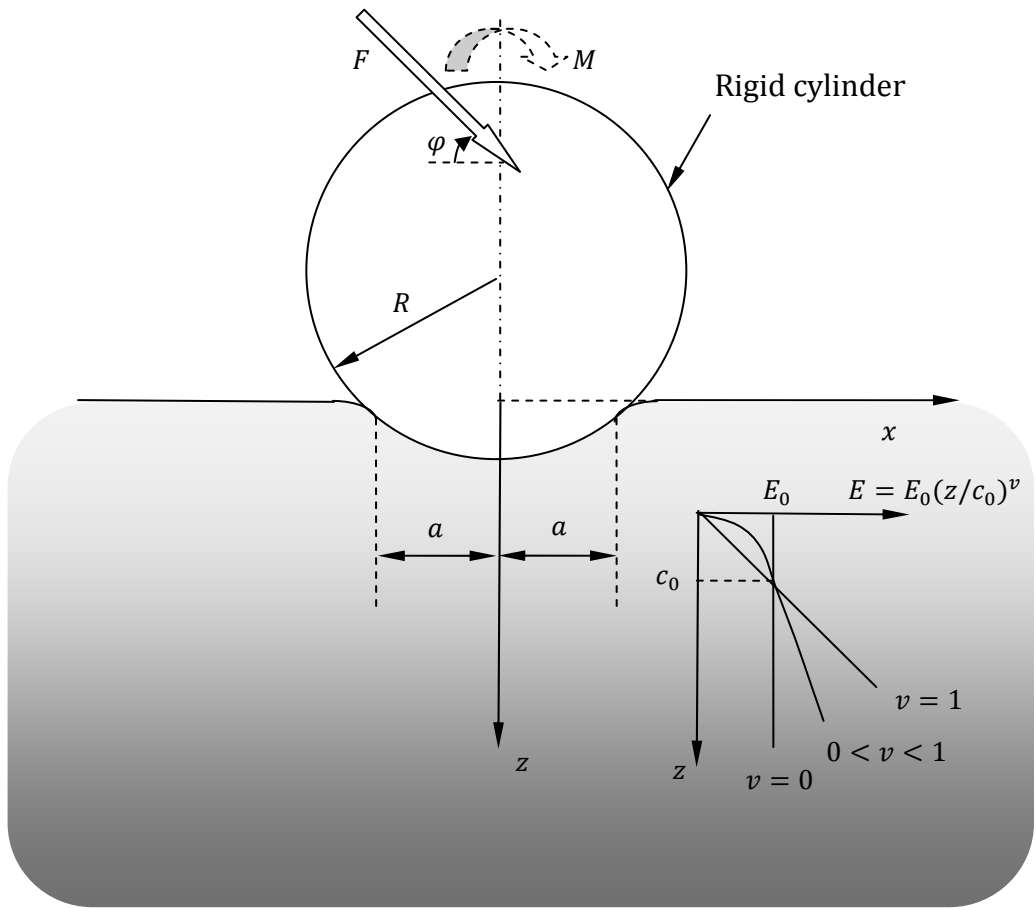

**Figure 1**

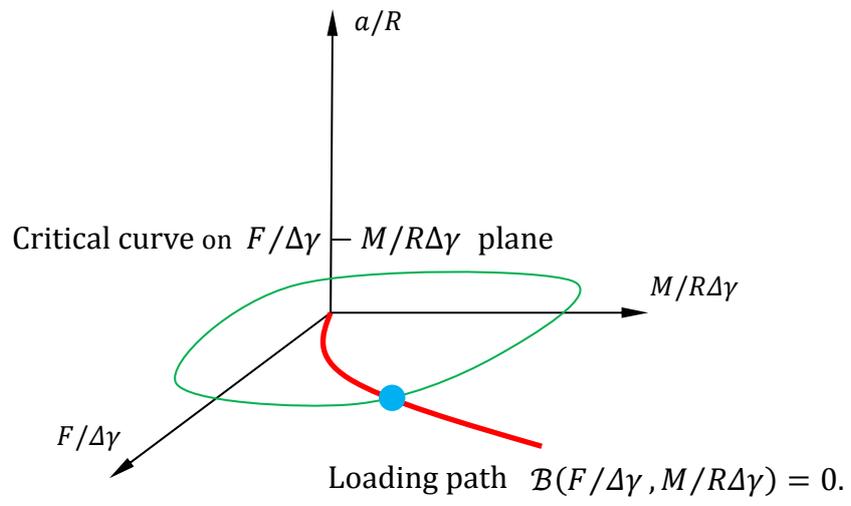

**Figure 2**

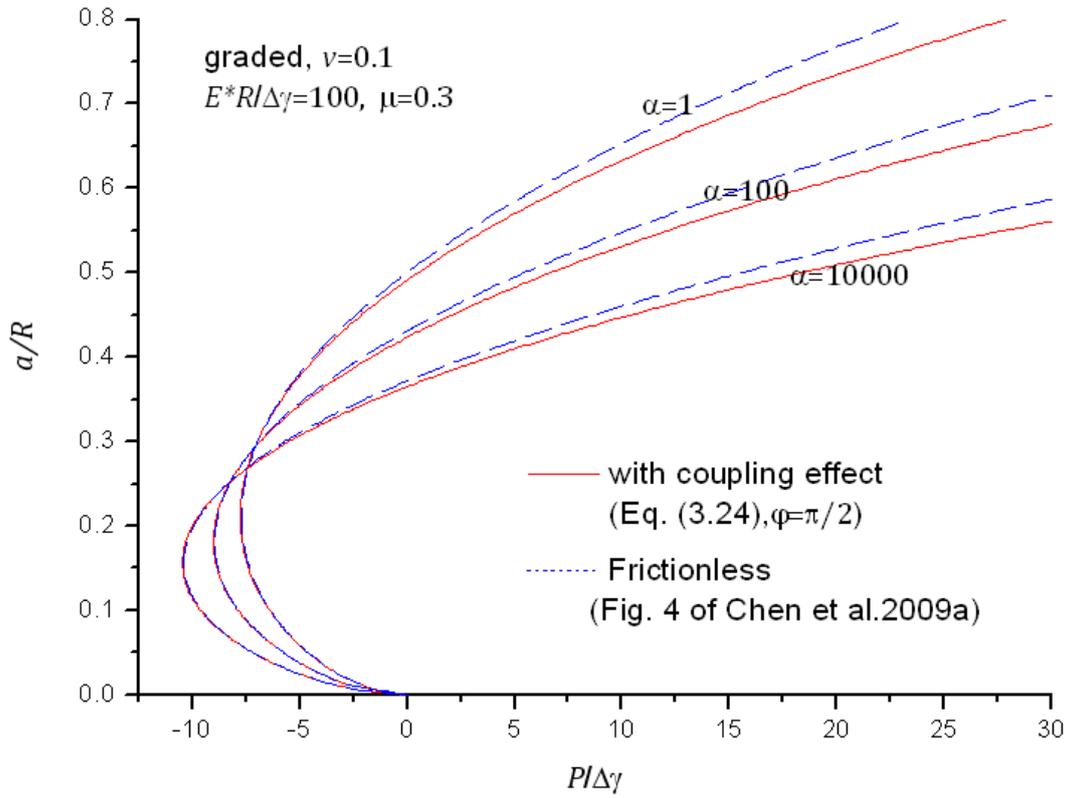

**Figure 3a**

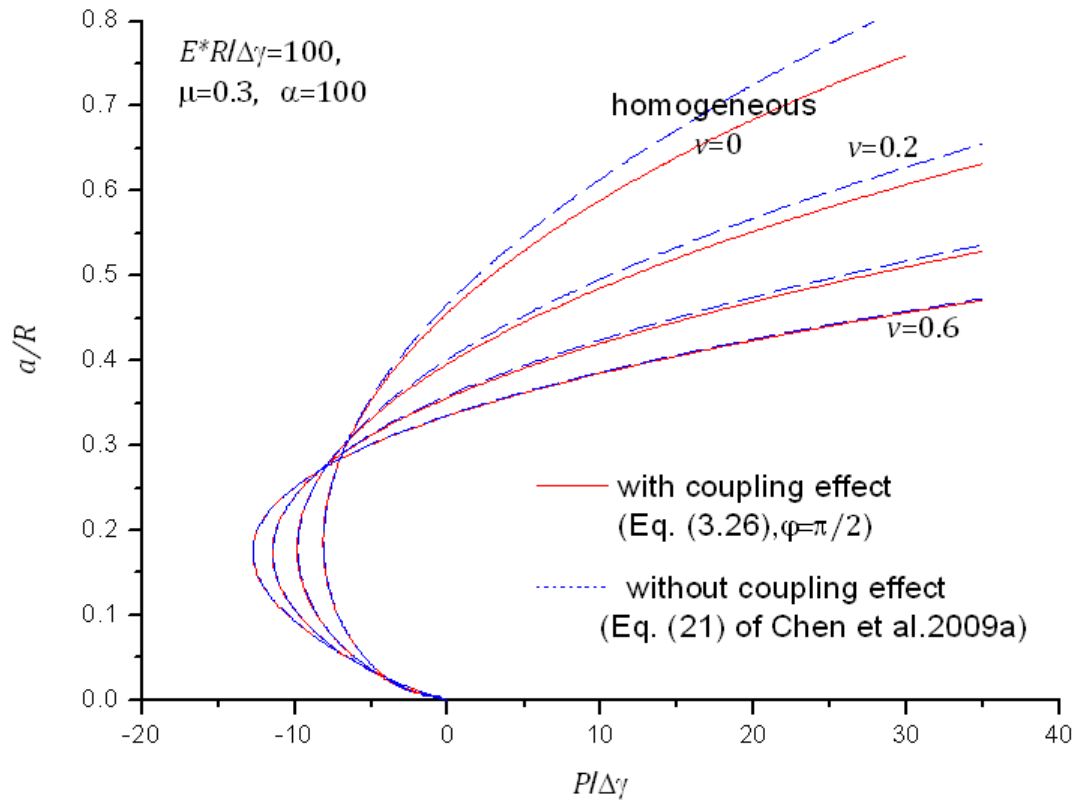

**Figure 3b**

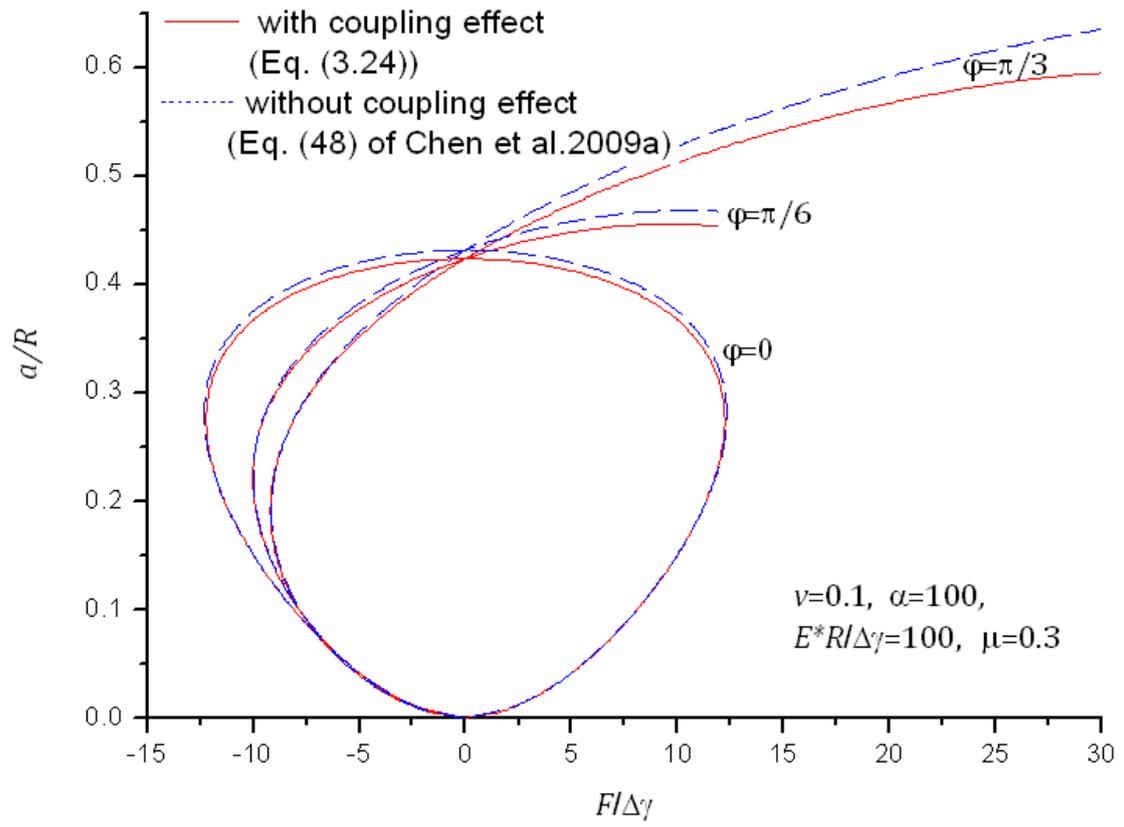

**Figure 4**

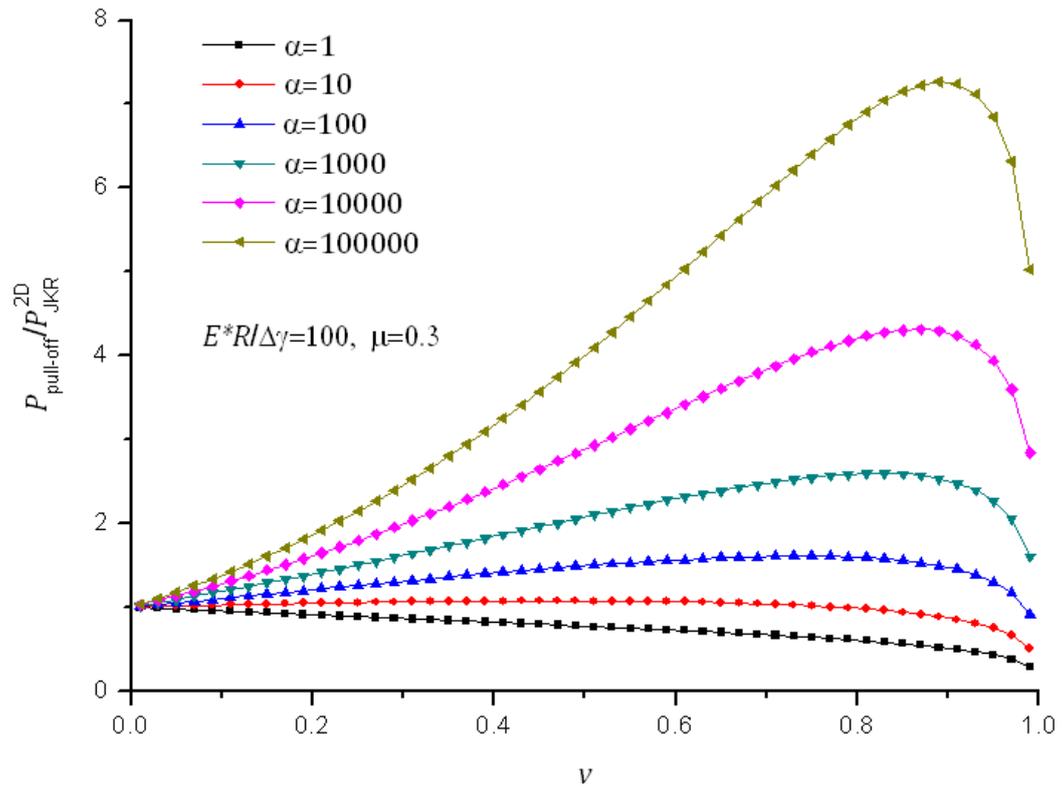

**Figure 5a**

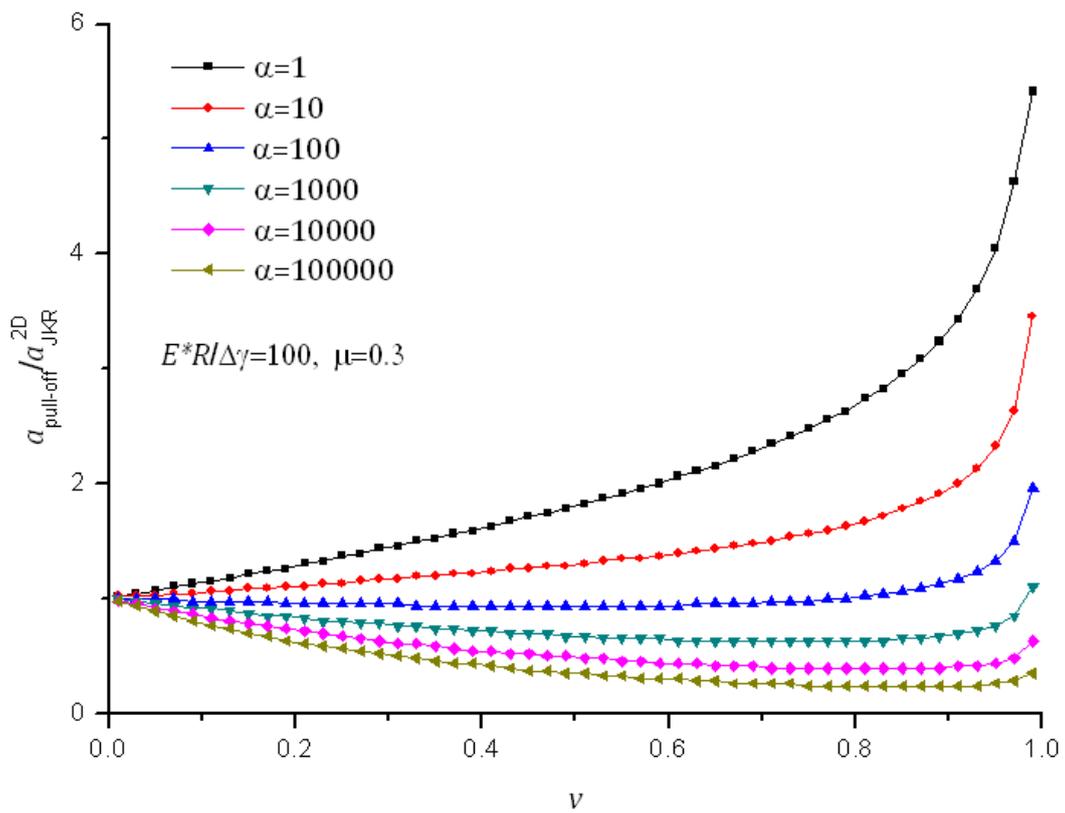

**Figure 5b**

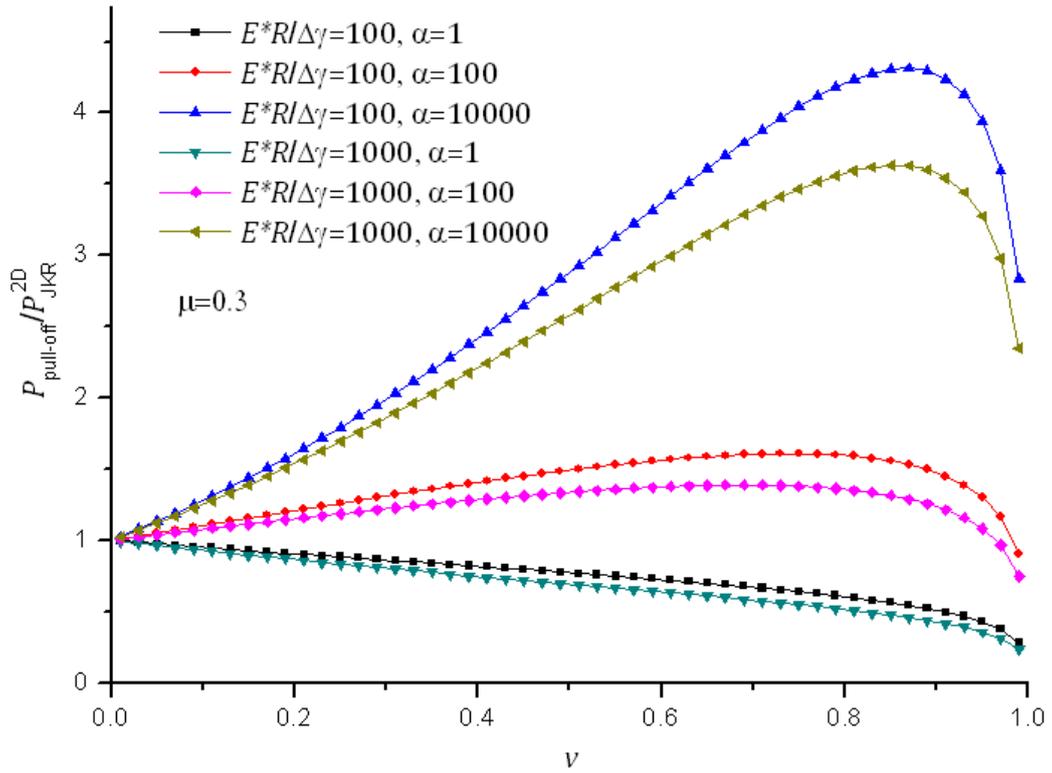

**Figure 6a**

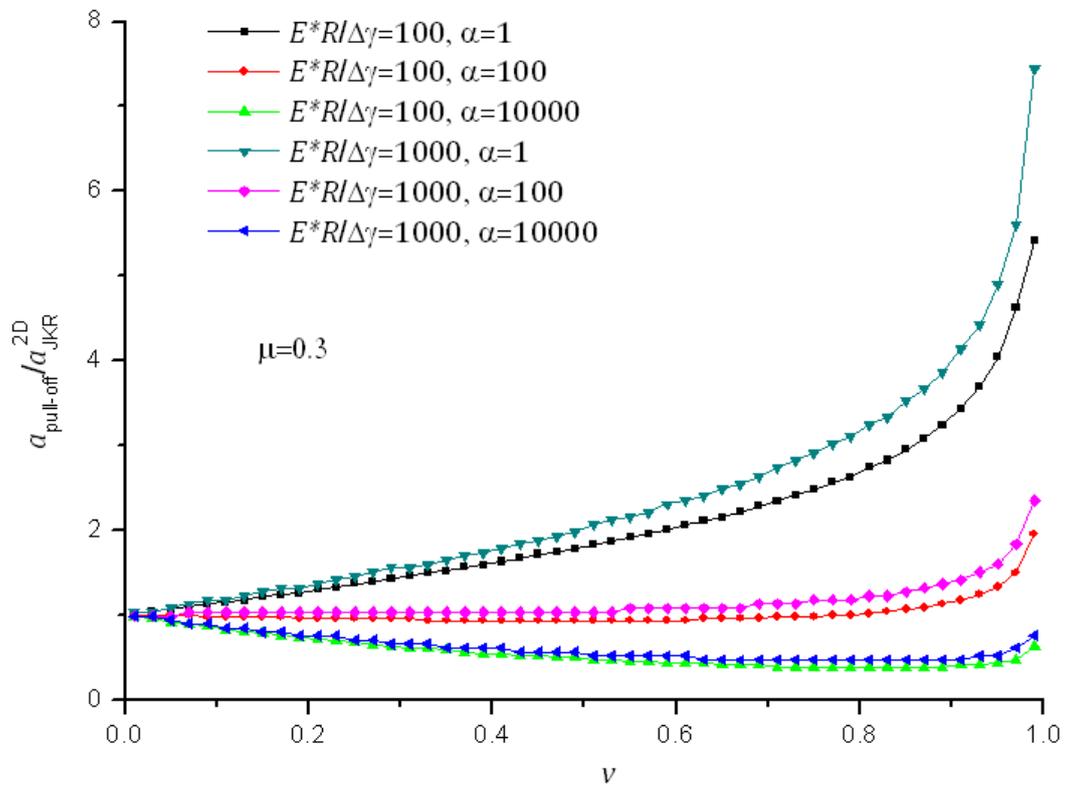

**Figure 6b**

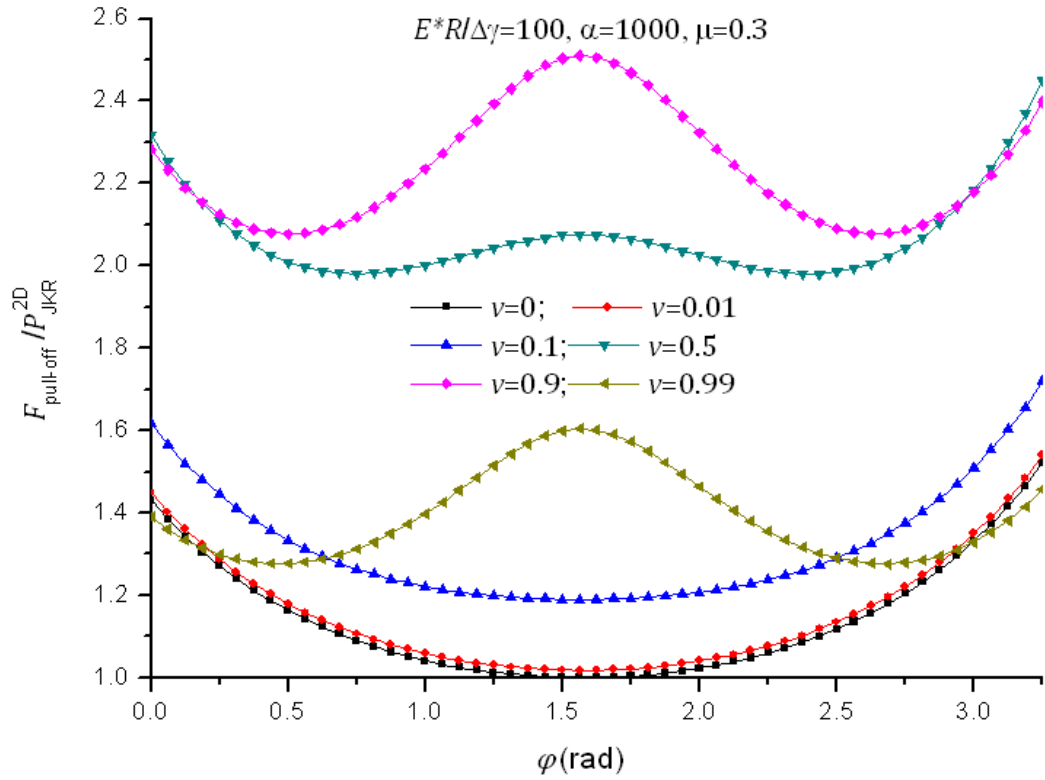

**Figure 7a**

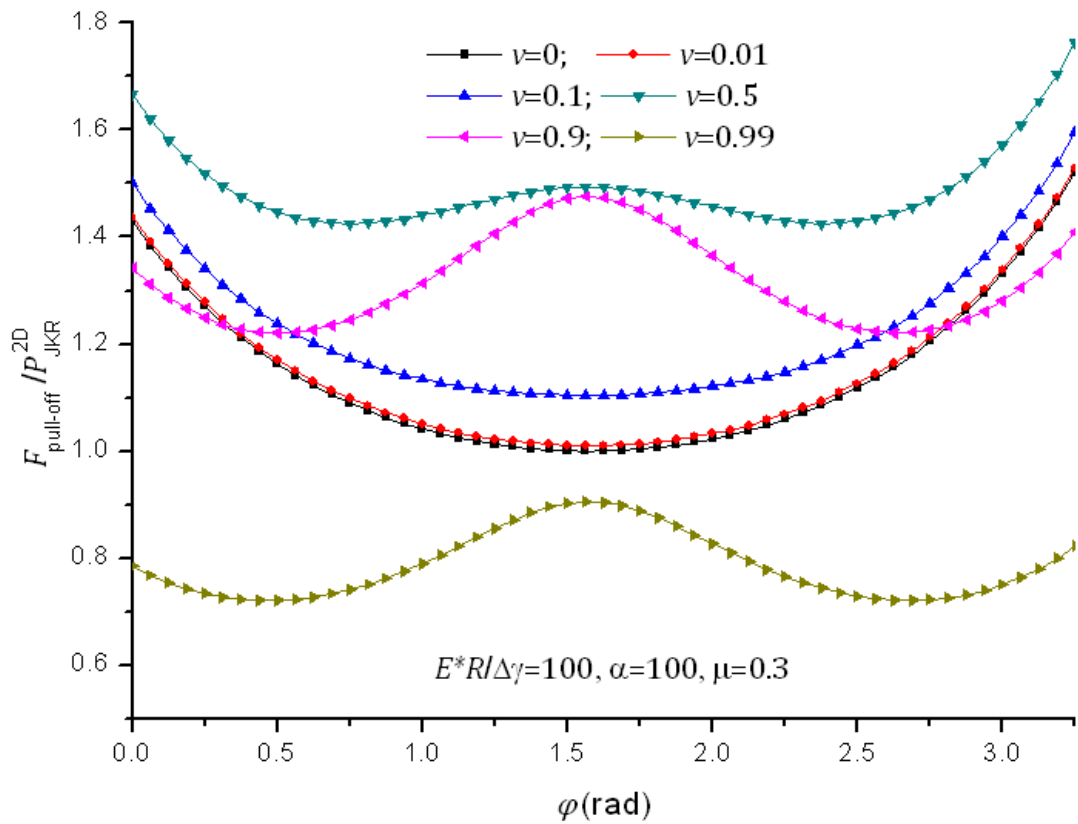

**Figure 7b**

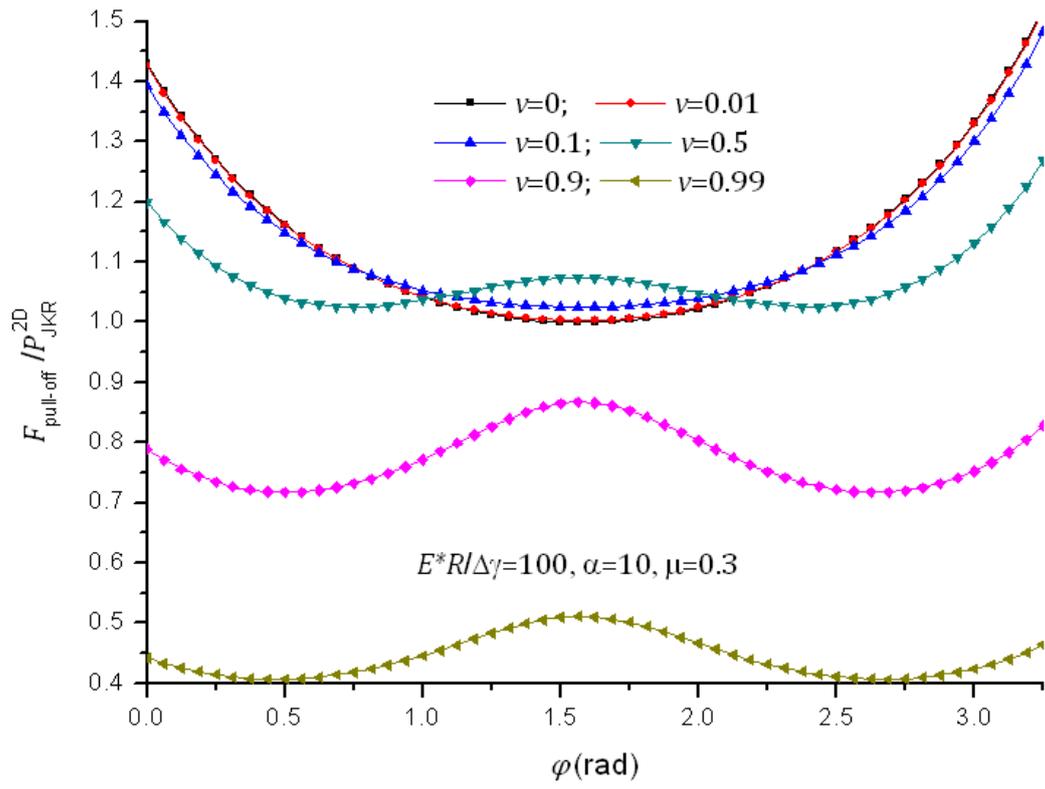

**Figure 7c**

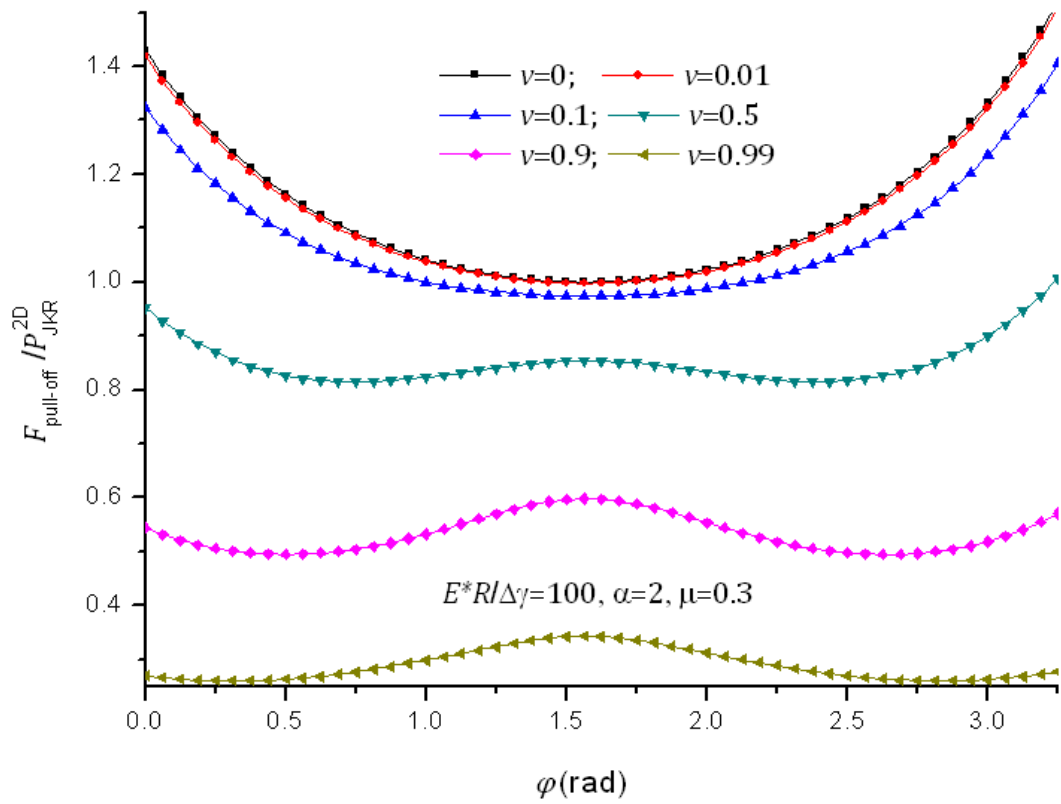

**Figure 7d**

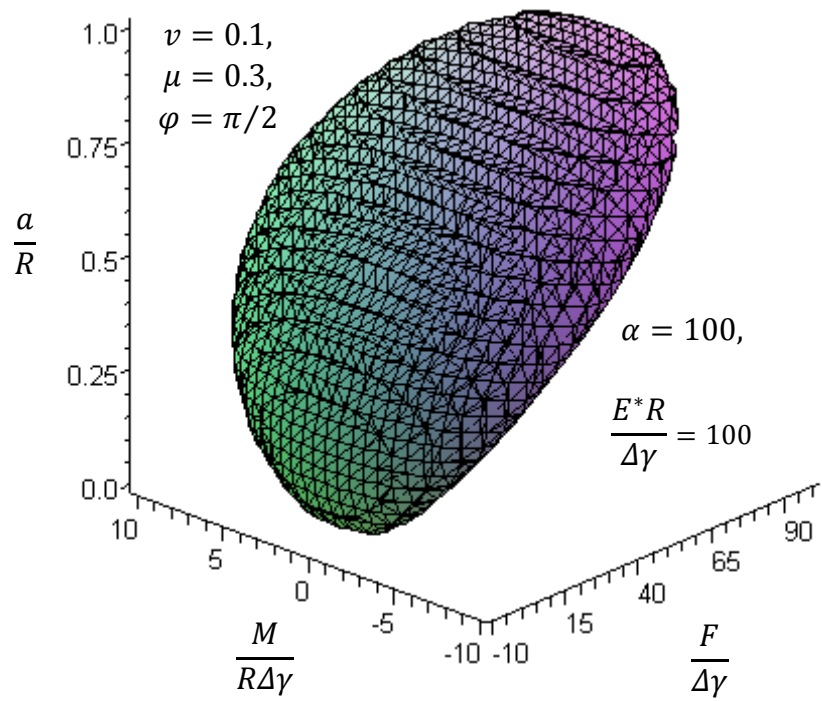

**Figure 8a**

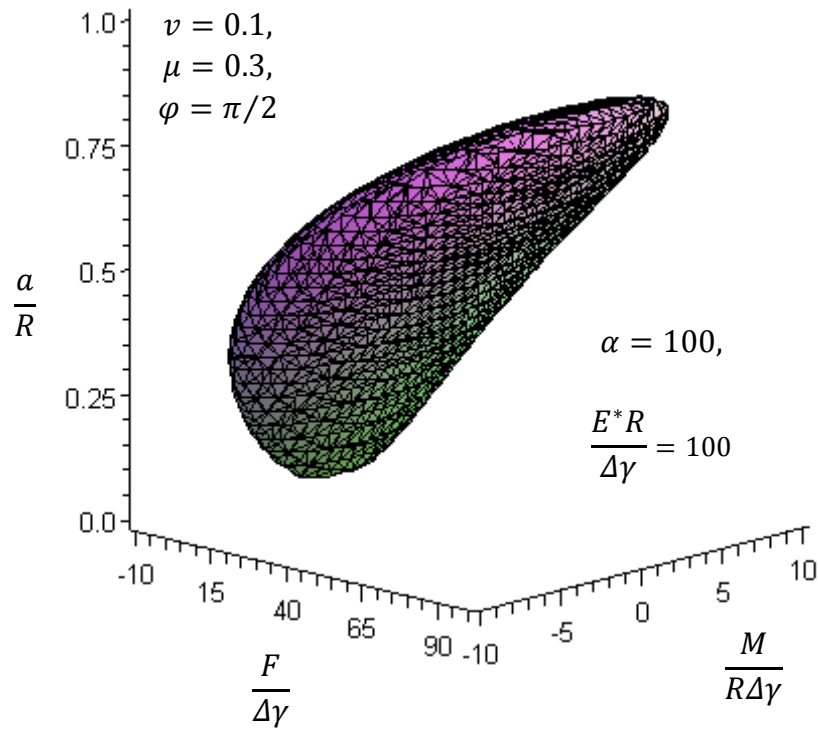

**Figure 8b**

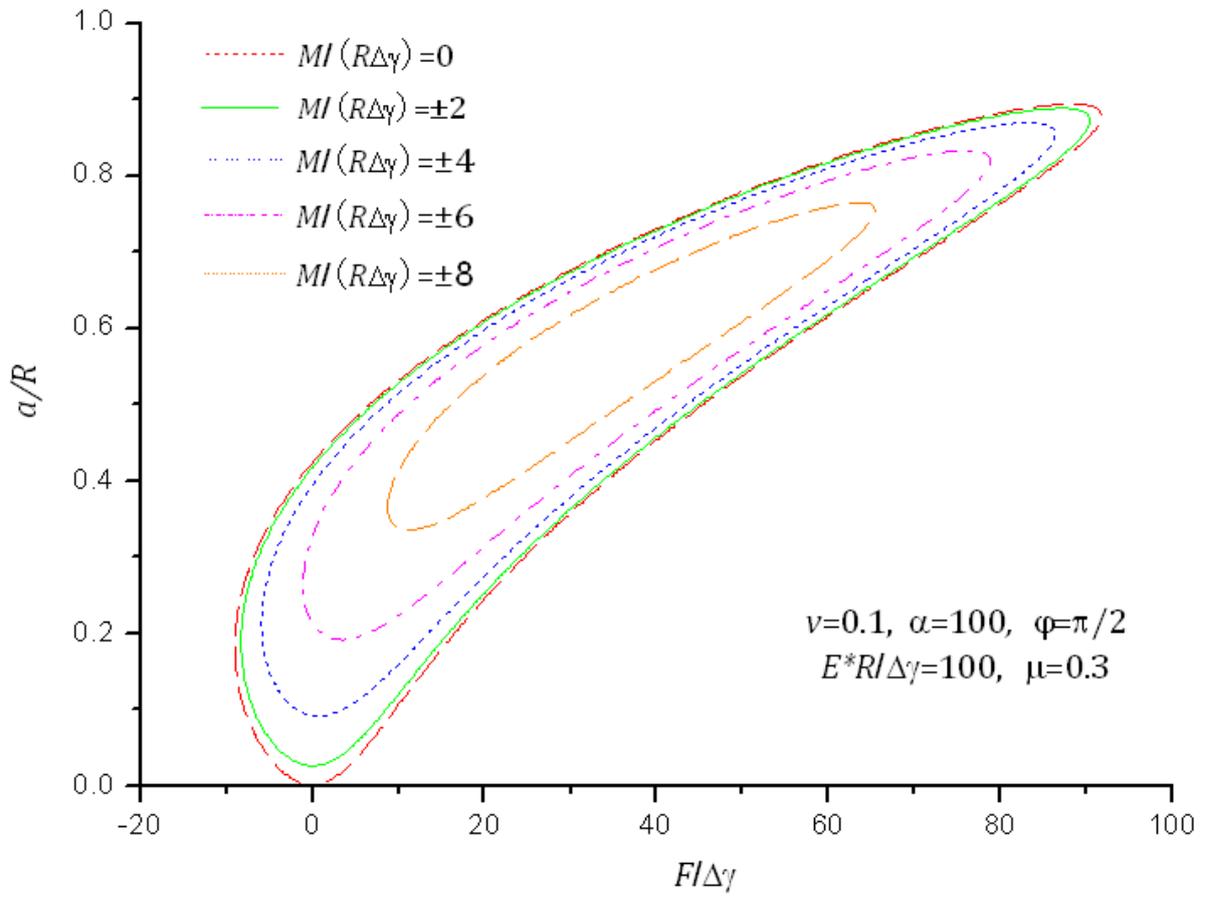

**Figure 8c**

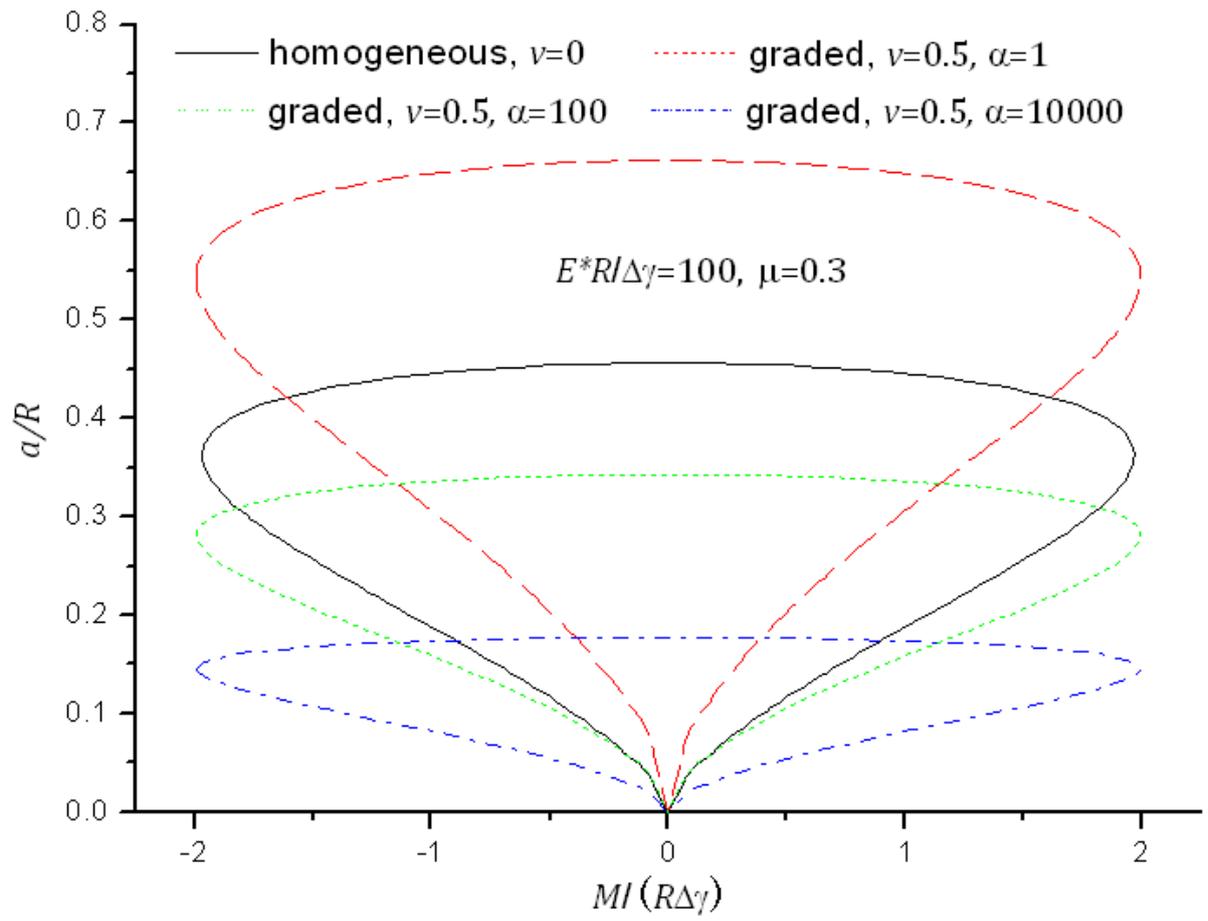

**Figure 9**

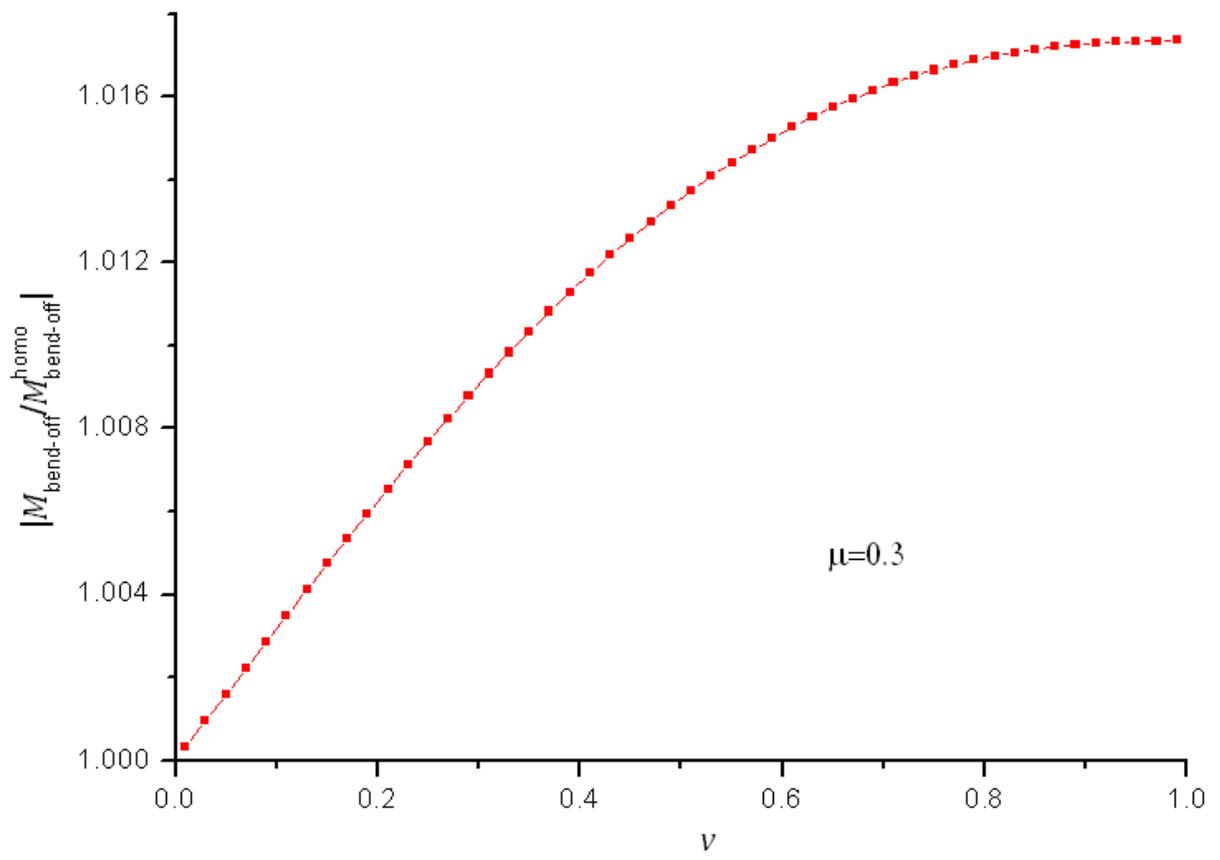

**Figure 10**

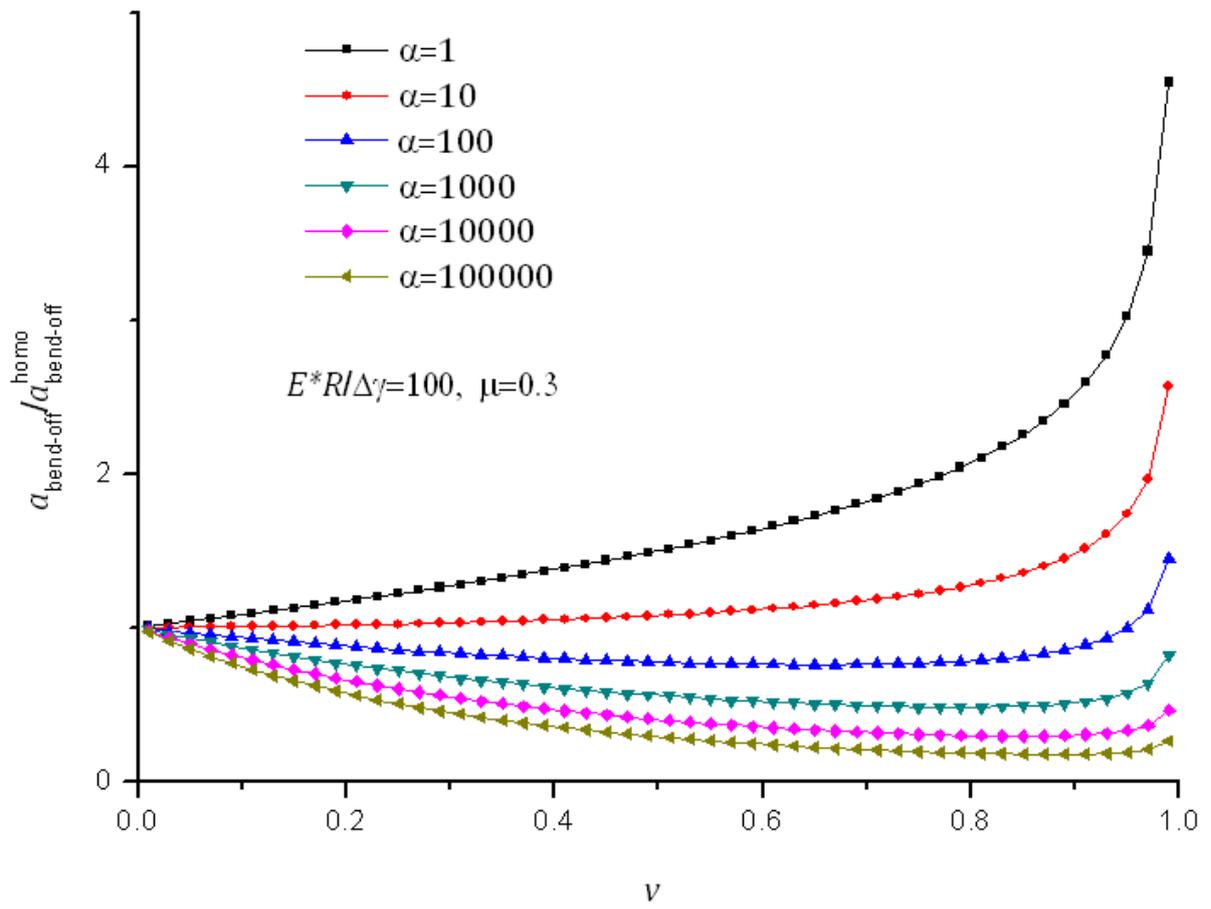

Figure 11

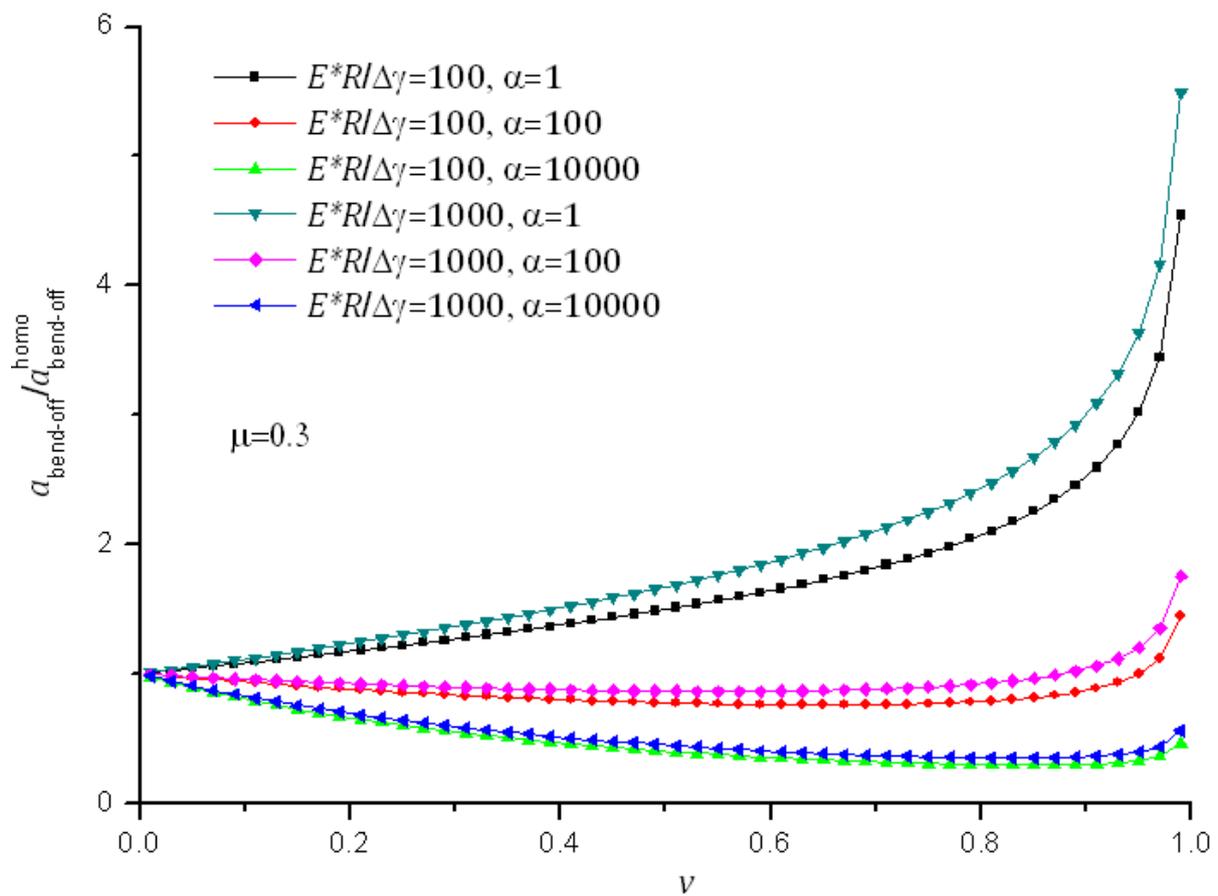

**Figure 12**